\documentclass[conference,review,anonymous]{IEEEtran} 
\pdfoutput=1 
\usepackage{cite}
\usepackage{amsmath,amssymb,amsfonts}
\usepackage{mathabx} 
\usepackage{tikz}
\usepackage{fbox}
\usetikzlibrary{positioning}
\usepackage{ragged2e}
\usepackage{hyperref}
\hypersetup{
    colorlinks=true,
    linkcolor=darkgray,
    filecolor=magenta,      
    urlcolor=gray,
    pdftitle={Overleaf Example},
    pdfpagemode=FullScreen,
    }

\newcommand{\hide}[1]{}

\usepackage{tcolorbox}
\usepackage{algorithmic}
\usepackage{graphicx}
\usepackage{textcomp}
\usepackage{url}
\usepackage{xcolor}
\usepackage{array, colortbl}
\usepackage{graphicx}
\usepackage[T1]{fontenc}
\usepackage{pifont}

\def\BibTeX{{\rm B\kern-.05em{\sc i\kern-.025em b}\kern-.08em
    T\kern-.1667em\lower.7ex\hbox{E}\kern-.125emX}}

\newcommand{\todo}[1]{\textbf{\textcolor{red}{#1}}}

\begin{document}

\definecolor{neutralBlue}{RGB}{62, 152, 179}
\definecolor{increaseRed}{RGB}{166, 41, 76}
\definecolor{decreaseGreen}{RGB}{41, 166, 102}
\definecolor{darkgrey}{RGB}{39, 39, 43}

\renewcommand{\upuparrows}{\textcolor{increaseRed}{\mathbin\uparrow\hspace{-.3em}\uparrow}}
\renewcommand{\downdownarrows}{\textcolor{decreaseGreen}{\mathbin\downarrow\hspace{-.3em}\downarrow}}
\newcommand{\down}{\textcolor{decreaseGreen}{\downarrow}}
\newcommand{\up}{\textcolor{increaseRed}{\uparrow}}
\newcommand{\no}{\textcolor{neutralBlue}{\bullet}}

\title{On the calibration of Just-in-time Defect Prediction}

\author{\IEEEauthorblockN{1\textsuperscript{st} Xhulja Shahini}
\IEEEauthorblockA{xhulja.shahini@paluno.uni-due.de}
\and
\IEEEauthorblockN{2\textsuperscript{nd} Jone Bartel}
\IEEEauthorblockA{jone.bartel@paluno.uni-due.de}
\and
\IEEEauthorblockN{3\textsuperscript{rd} Klaus Pohl}
\IEEEauthorblockA{klaus.pohl@paluno.uni-due.de}
\and
\IEEEauthorblockN{ 
\hspace*{28ex}
\textit{paluno-the Ruhr Institute for Software Technology} \\
\hspace*{28ex}
\textit{University of Duisburg Essen}\\
\hspace*{28ex} Essen, Germany \\}

}

\maketitle

\begin{abstract}
Just-in-time defect prediction (JIT DP) leverages machine learning to identify defect-prone code commits, enabling quality assurance (QA) teams to allocate resources more efficiently by focusing on commits that are most likely to contain defects.
Although JIT defect prediction techniques have introduced notable improvements in terms of predictive accuracy, they are still susceptible to misclassification errors such as false positives and false negatives. 
This can lead to wasted resources or undetected defects, a particularly critical concern when QA resources are limited.
To mitigate these challenges and 
preserve the practical utility of JIT defect prediction tools, it becomes essential to estimate the reliability of the predictions, i.e., computing confidence scores. 
Such scores can help practitioners 
determine the trustworthiness of predictions and
%identify predictions that are most likely to be correct.
and thus prioritize them efficiently. 

A simple approach to computing confidence scores is to extract, alongside each prediction, the corresponding prediction probabilities and use them as indicators of confidence.
However, for these probabilities to reliably serve as confidence scores, the predictive model must be well-calibrated.
%ensuring that it computes prediction probabilities that accurately reflect the likelihood that a prediction is correct. 
This means that the prediction probabilities must accurately represent the true likelihood of each prediction being correct.
Miscalibration, common in modern machine learning models, distorts probability scores such that the model’s prediction probabilities do not align with the actual probability of those predictions being correct. %hence leading to poor prioritization and resource allocation.
Despite its importance, model calibration has been largely overlooked in JIT defect prediction. 

In this study, we evaluate the calibration of several state-of-the-art JIT defect prediction techniques to determine whether and to what extent they exhibit
poor calibration. Furthermore, we assess whether post-calibration methods can improve the calibration of existing JIT defect prediction models.
Our experimental analysis reveals that all evaluated JIT DP models exhibit some level of miscalibration, with Expected Calibration Error (ECE) ranging from 2\% to 35\%. Furthermore, post-calibration methods do not consistently improve the calibration of these JIT DP models.
\end{abstract}


\section{Introduction}
In recent years, just-in-time defect prediction (JIT DP) has emerged as a valuable machine learning (ML)-based technique, designed to predict whether a code commit is defect-prone or clean. 
By identifying code commits that are more likely to contain defects, JIT defect prediction helps quality assurance (QA) practitioners decide whether to perform targeted inspections and code reviews, as well as where and how to allocate testing efforts and resources \cite{b3,b4}.
By supporting the prioritization of the code commits for further investigation and testing, JIT defect prediction models enable the timely identification of defects in the codebase. 
JIT defect prediction thus provides a means to optimize QA workflows.

Recent advances in JIT defect prediction have significantly enhanced predictive accuracy; however, these tools still produce a significant rate of false positive or false negative predictions\footnote{DeepJIT produces false positive rates of 26-39\%. CodeBERT4JIT produces false negative rates of 79-83\%.}. 
Trusting and acting on incorrect JIT defect prediction predictions might induce costly consequences.
For example, trusting and acting on a false positive, i.e., clean commits classified as defective, leads to wasting QA resources on carefully investigating code commits that contain no defects \cite{b7,b34}.
In contrast, trusting and acting on a false negative, i.e., defective commits classified as clean, poses the risk of overlooking defects, potentially allowing them to slip into production; 
thereby, inducing higher rework costs and reduced user satisfaction and system usage \cite{b1,b35}.
The consequences of misclassifications in JIT defect prediction become increasingly critical due to the limited availability of QA resources. 
Allocating resources to investigate clean commits means that fewer resources are available to assess commits that are likely defective.
Furthermore, as the codebase grows in size and complexity, available QA resources are only sufficient to thoroughly investigate and test a subset of the commits predicted to be defective.
This limitation highlights the importance
of computing reliable model confidence scores to help practitioners identify predictions that are most likely to be correct, and thus prioritize them efficiently. 
By doing so, teams can ensure that the commits most likely to contain defects (true positives) receive the necessary resources for in-depth investigation and testing.

Drawing upon the wider literature, including insights from the machine learning community \cite{b20,b22,b23, b29} medicine  \cite{b30,b31,b32}, and more recently the defect prediction community \cite{b5,b9}, a straightforward approach involves predicting not only the classification label but also the respective prediction probability as a measure of confidence on the correctness of this prediction. 
Confidence scores can serve as valuable indicators to aid practitioners in determining when to rely on a model's predictions. For example, a prediction made with 98\% confidence is likely a correct prediction, while one with 51\%\footnote{This example assumes a classification threshold 
T=0.5 (50\%), where predictions with confidence below T are classified as C=0, otherwise as C=1.} confidence is likely a misclassification. 
In addition, confidence scores can also be used further to rank potentially true positive predictions to facilitate their prioritization. 
This ranking approach has already been implemented in several defect prediction studies \cite{b3,b10,b11,b12}. A similar approach is also commonly used in effort-aware JIT defect prediction, where predictions are prioritized based on their estimated effort, calculated as the ratio of the prediction probability over the code complexity or the lines of code.

\emph{Problem:} Utilizing prediction probabilities to identify correct predictions requires that the predictive model be well calibrated, meaning that the predicted probabilities reflect the true model’s accuracy. 
For example, a calibrated defect prediction probability of 0.8 implies 80\% likelihood that a given commit is defective. 
However, the literature has shown that modern ML, and especially DL models, often suffer from poor calibration \cite{b17,b18}. 
Further studies have also reported that different ML algorithms display varying degrees of calibration\cite{b19,b20}. 
A poorly calibrated model generates prediction probabilities that might fail to accurately represent the true likelihood of correct predictions. 
Miscalibrated defect prediction probabilities can mislead practitioners on which predictions to trust and act upon, potentially leading to suboptimal or even inefficient prediction prioritization and QA resource allocation. 

The miscalibration of ML models and the importance of model calibration is a widely discussed topic in the broader literature. However, in the JIT defect prediction community, calibration assessments and miscalibration correction of predictive models remain a largely neglected topic, even though predicted probabilities are already being used by several techniques to rank and prioritize defect predictions. 

\emph{Contribution:}
Via this study, we address this gap by empirically evaluating the calibration level of several existing state-of-the-art JIT defect prediction models, to assess if they also exhibit poor calibration. This assessment can help determine whether the raw prediction probabilities can be used to reliably inform QA practitioners on whether to trust JIT defect predictions and how to prioritize them. 
Furthermore, we assess whether post-calibration methods can improve the calibration of existing JIT defect prediction techniques.
Post-calibration methods aim at correcting the calibration of a predictive model's prediction probabilities.
We focus on only post-calibration methods as they do not require either the modification of the predictive model design or the re-training of the predictive model. Our objective is not to introduce a novel, better-calibrated JIT defect prediction technique. Our goal is to assess whether existing ones exhibit miscalibration issues and, if so, assess whether we easily recalibrate them without having to modify the JIT DP model architecture. \\
Concretely, in this study, we make the following contributions:  \\
\textbf{\ding{182}} We experiment with three state-of-the-art JIT defect prediction techniques, each employing a different ML model as a predictor. We measure the extent to which they are miscalibrated and determine if JIT defect prediction techniques based on different ML models exhibit varying levels of calibration.\\
\textbf{\ding{183}} We conduct experiments with two different datasets to examine how different datasets may affect the calibration levels of these JIT defect prediction models.\\
\textbf{\ding{184}} Lastly, we assess the potential of post-calibration methods to improve the calibration levels of these JIT DP models. We experiment with two state-of-the-art post-calibration methods, as proposed in \cite{b17,b18}, to measure their effect on the calibration levels of the selected JIT DP models.

Through this empirical study, we address the following research questions:\\
\emph{\textbf{RQ1:} To what extent are existing JIT defect prediction models miscalibrated?} 

Our experimental results indicate that DeepJIT demonstrates an average ECE miscalibration rate of 35\% when trained on OPENSTACK data and 33\% when trained on QT data. In comparison, CodeBERT4JIT exhibits an average ECE miscalibration rate of 12\% with OPENSTACK data and 8\% with QT data. LApredict exhibits the lowest miscalibration rates, with an average of 9\% for OPENSTACK data and 3\% for QT data.\\
\emph{\textbf{RQ2:} To what extent can post-calibration methods improve the calibration levels of existing JIT defect prediction models?}

Our experimental results demonstrated that Platt scaling notably reduced miscalibration rates, particularly for JIT defect prediction models with the highest initial miscalibration. The most notable improvements were observed in DeepJIT and CodeBERT4JIT. However, its effect on LApredict was minimal. In contrast, Temperature scaling exhibited variable performance across different JIT defect prediction models.
The effects of both Platt and Temperature scaling on DeepJIT and CodeBERT4JIT were statistically significant across all measured metrics. However, for LApredict, their statistical significance was observed in only a few of the measured metrics.

\section{Calibration}

ML classifiers generate predictions in the form of prediction probabilities that indicate the likelihoods that a given input belongs to each of the class labels known from the training data.
In binary classification tasks, the probability scores are typically calculated using a sigmoid activation function.
For example, given a new code commit, a JIT DP model might output a probability of 0.8 for the\emph{	\rmfamily{defective}} class (C=1), indicating that the model believes, with 80\% probability, that the given input belongs to the class \emph{\rmfamily{defective}} and with 20\% probability (1 - 0.8) that the given input belongs to the class \emph{\rmfamily{clean}} (C=0). 
Based on a classification threshold, typically set to 0.5, the predicted class label will be the one with prediction probability higher or equal to the threshold. In this case, the predicted probability for the class \emph{\rmfamily{defective}} surpasses the threshold (0.8 $\geq$ 0.5) therefore the model would predict the commit to be \emph{\rmfamily{defective}}. 

In various fields, particularly those utilizing ML models to support decision-making, practitioners require the model to deliver a prediction probability in addition to the predicted label \cite{b17,b18,b39}. If well-calibrated, these probabilities aid practitioners decide wether to rely on a prediction or not.

Calibration is a performance metric that measures how well the prediction probabilities reflect the accuracy of that model's predictions.
A predictive model is said to be well-calibrated if the prediction probabilities match the accuracy of the model. 
For example, if a well-calibrated model makes 10 predictions with a probability of 0.9 for each prediction, then the model is expected to accurately classify (90\%) nine of the instances. \\
\emph{Formal definition:}\\
Given: \\
1) a labeled dataset of N sets of random variables \( D = \{(x_i, y_i)\}_{i=1}^N \), where \( x_i \) represents the input features and \( y_i \) $\in {0,1}$  represents the corresponding label; and\\
%, following a joint probability distribution $\pi (x_i, y_i ) = \pi(y_i |x_i) \pi (x_i)$ 
2) a predictive model \( f \) trained on \( D \), which  given \( x_i \) computes  \( f(x_i) =  \{(\hat{y}_i, \hat{p}_i)\} \), where  \( \hat{y}_i \) is the predicted label, and \( \hat{p}_i \) is the predicted probability representing the model's confidence on \( \hat{y}_i \) being the true label of \( x_{i} \).
%: \( \hat{p}(\hat{y}_i = y_i | x_i ) \) .\\
We consider \( f \) to be well-calibrated if, for a set of predictions \( \hat{Y} \) with predicted probabilities \( \hat{P} \), 
the proportion $p$ of correct predictions among all predictions
%\( p(\hat{y}=y | x ) \)  
also approximates \( \hat{P} \).
\begin{equation}
P(\hat{Y} = Y \mid \hat{P} = p) = p, \forall p \in [0, 1]
\end{equation}

\subsection{Calibration metrics}
\label{calibration_metrics}

Calibration metrics quantify the discrepancy between a model's predicted probabilities and the observed rate of the correct predictions.
The most commonly used calibration metrics in the literature are the reliability diagram, Expected Calibration Error (ECE), Maximum Calibration Error (MCE), and Brier score\footnote{Despite ECE, MCE, and Brier score being frequently referred to as calibration metrics, they actually measure the calibration error, i.e., the miscalibration of a model, where higher values reflect higher miscalibration.} \cite{b17, b18, b21}.
To compute the first three metrics, we must first group predictions, based on a predefined binning schema, as follows:\\
Given predefined number of bins \emph{B}$\in\mathbb{N}$, and a set of $M$ predictions \(  \{(\hat{y}_i, \hat{p}_i)\}_{i=1}^M \) made by \( f \):
\begin{enumerate}
    \item Partition the prediction probability interval [0,1]  into \emph{B} subintervals (bins) defined by the interval endpoints: \( I_b=(\frac{b -1}{B}, \frac{b}{B} ] \), for each \( b = \{ 1,2,...,B \} \).
    \item Assign all the predictions into bins such that each bin \( b \) contains all the predictions \emph{i} whose prediction probability for class C=1, i.e., \emph{\rmfamily{defective}}, \( \hat{p}_i \) falls within the interval of \( I_b \).
    \item Calculate the accuracy \( A_b \) for each bin by computing the ratio of the instances whose true label is \emph{\rmfamily{defective}}, out of all the instances in the bin: \( A_b= \frac{1}{|b|} \sum_{i \in b} 1(y_i = \hat{y}_i) \).
    \item Calculate the confidence of each bin \( C_b \) as the average of the prediction probabilities for C=1 of all instances in the bin: \( C_b= \frac{1}{|b|} \sum_{i \in b} \hat{p}_i \).
\end{enumerate}
Literature proposed two different schemas for constructing bins: a) equal width bins\cite{b17}, and b) equal frequency bins\cite{b21}. 
The equal width (equiwidth) binning schema splits the prediction probability interval [0,1] into B subintervals of the same width, $\frac{1}{B}$. The equal frequency binning schema (adaptive) splits the prediction probability interval [0,1] into B subintervals, such that each subinterval contains an equal number of predictions, i.e., the cardinality of each bin is$\sim$ $\frac{M}{B}$. 

\emph{Reliability Diagram:} serves as a visual tool for assessing model calibration, plotting for each bin, the bin’s confidence on the x-axis against its accuracy on the y-axis. In a perfectly calibrated model, the diagram would form a diagonal line (an identity function), where the model's confidence, i.e., prediction probabilities, perfectly align with the accuracy of the model. Although perfect calibration is practically unattainable, the goal is to achieve a reliability diagram that closely follows this diagonal. 
Reliability diagrams aid in identifying probability ranges where the model exhibits overconfidence (a bin's confidence exceed its accuracy) or underconfidence (a bin's accuracy exceeds its confidence).

\emph{Expected Calibration Error}: (ECE) is a metric that quantifies a model's overall calibration by measuring the average discrepancy between predicted probabilities and actual accuracy across all predictions. ECE is computed by calculating the weighted average of the difference between accuracy and confidence within each bin, for all the bins.
ECE provides an overall aggregated measure of model calibration. 
An ECE= 0 denotes a perfectly calibrated model. Conversely, an ECE=1 denotes a fully miscalibrated model.

\emph{Maximum Calibration Error:} (MCE) is a metric that quantifies the peak model miscalibration by identifying the largest observed disparity between predicted confidence and actual accuracy. 
MCE is calculated as the maximum difference between accuracy and confidence across all bins, highlighting inputs for which the model exhibits the most under/over-confidence. 
As with ECE, MCE values close to zero denote good calibration.

\emph{Brier score:} is a bin-independent metric for measuring a model's calibration. 
It is computed as the mean squared error between predicted probabilities and the respective actual class label.
A low Brier score indicates a  well-calibrated model.

\subsection{Calibration methods}
\label{calibration_methods}
Calibration methods are designed to adjust the predicted probabilities of machine learning models, aligning them more closely with the model's actual accuracy. These methods help correct the calibration of over/under-confident models.

\hide{
Model calibration can be achieved via two main approached: 
\begin{itemize} 
    \item Via \textbf{Design-time Calibration}. 
    This approach integrates the calibration process directly into model training, often by using specialized loss functions or regularization techniques that promote well-calibrated outputs \cite{b28}. 
    This approach encourages the model to learn to generate calibrated probability estimates throughout training. 
    Design-time calibration requires changes to the model architecture and training, potentially impacting overall model performance. \todo{remove???} 
    \item Via \textbf{Post-calibration}. In contrast, this approach applies calibration methods to an already-trained model to transform the model's output. Popular post-calibration techniques include Platt scaling, Temperature scaling, and isotonic regression \cite{b17, b18, b21}. These methods require a separate calibration data subset —distinct from the training set— to fit the calibration model. This step is necessary to ensure that the calibration remains unbiased. Using the training data for calibration can lead to overfitting, resulting in overly optimistic confidence measurements \cite{b17}.
\end{itemize}}

A commonly used approach to correct model miscalibration, is to employ post-calibration methods to adjust the outputs, i.e., predicted probabilities of
an already-trained model. State-of-the-art post-calibration methods include Platt scaling and Temperature scaling \cite{b17, b18, b21}. These methods require a separate calibration data subset -distinct from the training set- to fit the calibration model. This step is necessary to avoid calibration measurements bias. Using the training data for calibration can lead to overfitting, resulting in overly optimistic calibration measurements \cite{b17}.

\emph{Platt scaling \cite{b22}:} utilizes the raw outputs, of a model's predictions, specifically the prediction logits, to fit a logistic regression model that generates probability estimates more closely aligned with the model's accuracy. 
The Platt scaling method adjusts calibrated probabilities as follows: 
\begin{equation}
    p_{calibrated} = \sigma(\alpha*q + \beta)
\end{equation}

where $\alpha$, $\beta$ $\in$ $\mathbf{R}$ represent the logistic curve parameters, $q$ represents the prediction logits derived from the model predictions, and $\sigma$ represents a sigmoid function. 
The optimization of $\alpha$ and $\beta$ is achieved by minimizing the Negative Log Likelihood (NLL) loss on the calibration set. 
This process results in better-calibrated probabilities that follow the shape of a logistic curve in a reliability diagram. 

\emph{Temperature scaling \cite{b17}:} extends Platt scaling by using only a single parameter, $T$, which scales the prediction logits through division:
\begin{equation}
    \alpha = 1/T, \beta = 0
\end{equation}
The optimal temperature $T$ is identified by minimizing the NLL loss specific to $T$ on a calibration set. When $T>1$, the difference between the output logits decreases, resulting in more evenly distributed calibrated probabilities. 

Both Platt and Temperature scaling adjust prediction probabilities without altering the predictive model's accuracy. 

\hide{
\emph{Isotonic Regression} adjusts predictions by learning an optimal monotonic transformation via a non-parametric function
f that maps uncalibrated predictions to better-calibrated ones:
\begin{displaymath}
    p_{calibrated} = f(p)
\end{displaymath} 
where p represents the uncalibrated prediction probability. The piecewise constant function f minimizes the squared loss
\begin{math}
    \sum_{i=1}^{n}(f(p_{i}) - y_{i})^{2}
\end{math}
over a calibration set, where $y_{i}$ denotes true labels associated with probabilities $p_{i}$.
}

\section{STUDY DESIGN}
Below we present the design of our study, including the evaluated JIT defect prediction techniques and the datasets.

\subsection{Subject Systems}

In this study, we evaluate three widely recognized JIT defect prediction techniques: DeepJIT \cite{b24}, LAPredict \cite{b25}, and CodeBERT4JIT \cite{b26}. We selected these techniques based on the fact that they are published in top-tier venues and have a significant citation count (DeepJIT: 249, LAPredict: 87, CodeBERT4JIT: 91), as indexed on Google Scholar\footnote{\url{https://scholar.google.com/}}. 
Moreover, each of these techniques utilizes a distinct ML model to train a defect predictor, enabling the assessment and comparison of the calibration across various ML models used for defect prediction. 
Additionally, all three JIT defect prediction techniques have been originally evaluated on the same datasets, namely QT and OPENSTACK, which enables a fair comparison of the calibration levels of different JIT defect predictors on a common data benchmark. 

\textbf{DeepJIT:} is a JIT defect prediction model that uses deep learning (CNN) to automatically extract semantic features from code commit features rather than manually extract them.
The model takes as input commit features and commit messages to train a just-in-time defect predictor \cite{b24}.

\textbf{LApredict:} is a logistic regression (LR)-based defect prediction model. LApredict used code commits features, focusing on the added lines of code, referred to as the "la"-feature on the paper, to train a just-in-time defect predictor \cite{b25}.

\textbf{CodeBERT4JIT:} is a CodeBERT based JIT defect prediction model. CodeBERT in itself is a pre-trained natural language processing (NLP) model, based on a multi-layer bidirectional Transformer architecture. CodeBERT encodes both code lines and natural language descriptions into a single vector containing the semantic meaning of the input sequence. 
The CodeBERT4JIT model takes as input commit features and commit messages to train a just-in-time defect predictor \cite{b26}.

\subsection{Subject Datasets}

To evaluate the selected techniques, we utilize the same datasets referenced in the original studies of the chosen JIT defect prediction techniques. This selection aligns with the objective of our study, which is to evaluate existing JIT DP techniques from a calibration perspective rather than focusing solely on accuracy. Therefore, to minimize validity risks, we replicated the original studies by strictly adhering to the implementation details provided in the corresponding repositories, including model configurations, hyperparameters, and datasets.

Concretely, we use the QT and OPENSTACK datasets, originally compiled by McIntosh and Kamei \cite{b27}. 
Moreover, these datasets have become standard in Just-in-Time defect prediction research. 
Both datasets consist of diverse code commit artifacts, encompassing metrics describing the commit size, diffusion, history, and authorship.

\textbf{QT Dataset} is derived from the code commits of QT\footnote{https://www.qt.io/} project,
consisting of 17 different metrics, such as the number of lines added, the number of lines deleted, the number of files changed, code commit complexity, developer experience and the number of past defects in the files being changed. \\
\emph{Class Distribution:} The QT dataset contains 25150 commits, 8\% of them labeled as \emph{\rmfamily{defective}}. 
The dataset is pre-split into training (22579 instances) and test set (2571 instances).

\textbf{OPENSTACK (OS) Dataset} is derived from the OPENSTACK project code commits\footnote{ https://www.openstack.org/}.
The same as QT, the OPENSTACK dataset consists of 17 different code commit metrics.\\
\emph{Class Distribution:} The OS dataset contains 12374 commits 13\% of them labeled as \emph{\rmfamily{defective}}. The dataset is pre-split into training (11043 instances) and test set (1331 instances).\\
The full details of these datasets can be found at \cite{b27}.

\subsection{Methodology}

\subsubsection*{\textbf{RQ1: To what extent are existing JIT defect prediction models miscalibrated?}}
To address this research question, we perform the following procedure for each experimental configuration, i.e., JIT defect prediction technique and dataset:

\begin{enumerate}
    \item Initially, we shuffle and then split the training dataset in K=10 folds, using a cross-validation method.
    \item We allocate nine folds for training the JIT DP model, and utilize the remaining fold as a validation set.
    \item The trained model is then utilized to make predictions on the validation set instances. These predictions are then used to measure the accuracy and calibration of the trained model (see section \ref{dependent_variables} for the metrics).
    \item Steps 2 and 3 are repeated a total of ten times, following the cross-validation method, so that each of the 10 folds serves once as a validation fold.
    \item We select the most accurate predictive model among the 10 trained JIT DP models to make predictions on the test set instances. Based on these predictions we compute once again the accuracy and calibration metrics.
\end{enumerate}

This procedure is repeated 10 times for each experimental configuration to mitigate selection bias and to enable the evaluation of statistical significance. The combination of 10 repetitions and 
K=10 cross-validation results in a total of 110 measurements for each experimental configuration, comprising 100 evaluations on the validation sets (10x10) and 10 on the test set.

\subsubsection*{\textbf{RQ2: To what extent can post-calibration methods improve the calibration levels of existing JIT DP models?}}

In order to conduct the experiments required to address this research question, we perform the following procedure for each experimental configuration, i.e., each JIT DP model, dataset, and post-calibration method:
\begin{enumerate}
    \item We shuffle and then split the training dataset via cross-validation into K=10 folds.
    \item We allocate eight folds for training the JIT DP model. We reserve one fold as a calibration set, and utilize the last fold as a validation set \footnote{We use separate calibration and validation sets, to avoid biased calibration measurements due to overfitting.}.
    \item We use the trained JIT DP model to make predictions on the calibration set instances. These predictions are subsequently used to fit one of the subject post-calibration methods, which is then employed to calibrate the JIT DP model's predictions. 
    \item We use the trained JIT DP model and the fitted post-calibration method, to make predictions for instances in the validation set. These predictions are then used to compute the accuracy and calibration metrics, as computed in RQ1. 
    \item Steps 2-4 are repeated ten times in total following the cross-validation method such that each of the 10 folds serves once as a validation and once as calibration fold.
    \item  Among the 10 trained JIT DP models, the model with the highest predictive accuracy is selected. This model, together with its corresponding fitted post-calibration method, is used to generate predictions for the instances in the test set. Based on these predictions we compute once again the accuracy and calibration metrics.
    \item We then assess the effectiveness of the calibration methods by calculating the difference in calibration metric measurements prior to (RQ1 measurements) and after the model calibration (RQ2 measurements).
\end{enumerate}

Same as in RQ1, this procedure is repeated 10 times for each experimental configuration.
The measurements obtained from both experiments (RQ1 $\&$ RQ2) are analyzed to assess the statistical significance of our findings. Specifically, we evaluate the statistical significance of the observed impact of the calibration methods on the miscalibration levels of the selected JIT defect prediction techniques. The details of the statistical tests conducted in this study are provided in Section \ref{stat-sign}.

\subsection{Post-calibration methods}

In this study, we use Platt scaling and Temperature scaling as calibration methods. We select these two methods because they are widely recognized in the literature as state-of-the-art for model calibration \cite{b17, b18, b21}. Additionally, unlike other calibration methods, Platt scaling and Temperature scaling do not alter the model’s prediction accuracy. Guo et al. demonstrate that certain calibration methods can negatively impact prediction accuracy \cite{b17}.

\hide{only apply post-calibration methods to address calibration in pre-trained prediction models. Our aim is not to introduce a novel, better-calibrated JIT defect prediction approach; instead, we evaluate whether current techniques are sufficiently calibrated. If not, we examine whether they can be calibrated efficiently without altering their model structure or training process.
The initial experiments involved the use of Platt scaling, Temperature scaling, and Isotonic regression, as these are among the most widely adopted calibration methods. However, our experiments on calibrating CodeBERT with Isotonic regression demonstrated lower effectiveness compared to Platt and Temperature scaling, particularly with non-binning-based calibration metrics. 
Additionally, unlike the other two methods, Isotonic regression adversely impacted the prediction accuracy of the JIT DP technique, leading to a reduced AUC score. 
Detailed results are available in our repository \todo{Add Repo link}.
For the subsequent experiments, we therefore decided to .
The aim is to find the optimal parameters ($\alpha$ and $\beta$) for the logistic curve which minimize the negative log-likelihood (NLL) loss.
These parameters are then used to rescale the output logits, thereby generating better-calibrated prediction probabilities. 
To apply Temperature scaling, we follow the same procedure as with Platt scaling. However, in this case, we aim to find the optimal value of a single parameter: the temperature.}

The application of Platt and Temperature scaling methods adheres to the descriptions outlined in Section \ref{calibration_methods}. We use the predicted logits from the calibration set instances to fit the logistic regression models, a separate model for each post-calibration method. Subsequently, the JIT DP model is applied in conjunction with each trained post-calibration model individually to generate predictions for the validation and test set instances.
We utilize the open-source implementation of \href{https://github.com/EFS-OpenSource/calibration-framework/blob/main/netcal/scaling/LogisticCalibration.py}{Platt scaling} and \href{https://github.com/EFS-OpenSource/calibration-framework/blob/main/netcal/scaling/TemperatureScaling.py}{Temperature scaling} with default hyper-parameters, as provided by the official netcal repository.

\subsection{Evaluation Design}

\subsubsection{The \textbf{INDEPENDENT VARIABLES} of this study are}

\emph{Datasets:} {QT, OPENSTACK}.\\
\emph{JIT DP techniques:} {DeepJIT, LApredict, CodeBERT4JIT}.\\
\emph{Calibration methods:} {No calibration, Platt scaling, Temperature scaling.\\

\subsubsection{The \textbf{DEPENDENT VARIABLES} under study are}
\label{dependent_variables}

\textsc{Calibration metrics:}
To assess the calibration of the selected JIT DP techniques, we use the Expected Calibration Error (ECE), Maximum Calibration Error (MCE), Brier score and reliability diagrams. These metrics are widely recognized in the literature as standard calibration metrics \cite{b17, b18, b21}.

The computation of these metrics adheres to the descriptions outlined in Section \ref{calibration_metrics}. For implementation, we utilize the open-source resources provided by the official repositories of sklearn and netcal. Specifically, the implementations referenced are as follows: netcal metrics
\href{https://github.com/EFS-OpenSource/calibration-framework/blob/main/netcal/metrics/confidence/ECE.py}{ECE} and \href{https://github.com/EFS-OpenSource/calibration-framework/blob/main/netcal/metrics/confidence/MCE.py}{MCE}, and scikit-learn \href{https://scikit-learn.org/stable/modules/generated/sklearn.metrics.brier_score_loss.html#sklearn.metrics.brier_score_loss}{Brier Score} and  \href{https://github.com/scikit-learn/scikit-learn/blob/5491dc695/sklearn/calibration.py#L925}{Reliability Diagram}.

\emph{Parametrization:}
As outlined in section \ref{calibration_metrics}, computing calibration metrics (ECE, MCE, and reliability diagrams) requires the predefinition of the binning configuration, namely, \emph{B}: representing the number of bins, and the binning schema, which dictates the process for creating bins. \\
\underline{Number of bins:} To determine appropriate number of bins (B), we reviewed relevant literature on calibration measurement. Based on this review, we opted to experiment with two settings, specifically B={15, 50}. Due to the small sample size in the validation set, we refrain from using a larger number of bins to avoid too many empty bins or bins with few predictions, which could result distort the calibration scores.\\
\underline{Binning schema:} 
We examine both binning schemas commonly suggested in the literature: the \emph{equiwidth} and the \emph{adaptive} schema. This enables us to assess whether the JIT DP models generate predictions with evenly distributed probabilities across different bins, or if they show a tendency to cluster within certain probability ranges. 
%The former which may suggest overcalibration or undercalibration issues.

\subsubsection{\textbf{Statistical tests}}
\label{stat-sign}
To measure the statistical significance of the outcomes of this study we utilize the Wilcoxon Signed-Rank Test and the Paired t-test.
%We based our decision about the statistical tests on the following factors:\\
%\textit{Data Dependency:} Since we are comparing the same set of predictions before and after calibrating the predictive model, the data points are dependent, because the calibration metrics before and after calibration are computed on the same set of predictions from the defect predictor. Therefore, we should utilize a statistical test for paired data, rather than for independent samples. \\
%\textit{Metric Type:} The calibration metrics that we have computed in this study are continuous values, so the statistical test should be able to handle continuous data.\\
%\textit{Distribution Assumptions:} 
%To choose the right test for paired continuous data, we should first assess whether the metric values follow a normal distribution. This helps us determine if we should use a parametric or non-parametric statistical test. 

To test for normality, we apply \href{https://docs.scipy.org/doc/scipy/reference/generated/scipy.stats.monte_carlo_test.html#scipy.stats.monte_carlo_test}{Monte Carlo test},
 when the sample size is less than 50. Specifically, we use Monte Carlo test to assess the normality of the test subset samples.  For larger samples, i.e., validation folds samples, we use the \href{https://docs.scipy.org/doc/scipy/reference/generated/scipy.stats.normaltest.html}{D'Agostino \& Pearson test}. In both cases, statistical significance is set to a p-value threshold of 0.05. If the resulting p-value is below 0.05, we reject the null hypothesis, indicating that the data is not normally distributed.
%We evaluate the normality of each measured metric, namely $ECE_{Equiwidth}, ECE_{Adaptive}, MCE_{Equiwidth}, MCE_{Adaptive}$, and $Brier$. For each metric, we compare the distribution of measurements from the uncalibrated model against those obtained from the calibrated model, considering Platt scaling and Temperature scaling separately.
Based on the normality assessment, we select the appropriate statistical test. If the metric values are normally distributed, we employ the parametric Paired t-test. Otherwise, we use the non-parametric Wilcoxon Signed-Rank Test.

We then evaluate the statistical significance, via the Wilcoxon Signed-Rank Test or the Paired t-test, of Platt scaling's impact on the calibration of selected JIT defect prediction techniques by comparing each of the calibration metrics before and after applying Platt scaling. Similarly, we assess the effect of Temperature scaling by comparing the calibration metrics of the original, uncalibrated techniques with those obtained after applying Temperature scaling. In both cases, statistical significance is determined using a p-value threshold of 0.05. 

\hide{
\textsc{Accuracy metrics:} 
To assess the performance of the selected JIT DP techniques in terms of accuracy, we use the same accuracy metrics reported in the original studies. Additionally, each of the three JIT DP techniques is evaluated with the following threshold-dependent and threshold-independent metrics. For the threshold-dependent assessment, we compute the confusion matrix (TP, FP, TN, FN)  using a threshold of 
\emph{(t=0.5)}. This matrix enables the derivation of additional compound metrics, including accuracy, precision, recall, and F-measure. For the threshold-independent metric, we compute the Area Under the Curve (AUC) score, offering an aggregated measure of model accuracy across varying thresholds.\\
}

\section{Results}

In this section we present our experimental results and answer the research questions.

\subsection{RQ1: To what extent are existing JIT defect prediction models miscalibrated?}

\begin{table}[h!]
    \caption{DeepJIT Calibration Scores}
    \centering    
    \renewcommand{\arraystretch}{1.2}
    \setlength{\arrayrulewidth}{0.1mm}
    \setlength{\tabcolsep}{3pt}
    \begin{tabular}{|c|c|c|c|c|c|}
        \hline
        \rowcolor[HTML]{EFEFEF} 
        \textbf{Data subset} & \multicolumn{2}{c|}{\textbf{ECE \%}} & \multicolumn{2}{c|}{\textbf{MCE \%}} & \textbf{Brier \%} \\
        & Equiwidth & Adaptive & Equiwidth & Adaptive & \\
        \rowcolor[HTML]{EFEFEF} 
        & 15 bins & 15 bins & 15 bins & 15 bins &  \\
        \hline
        \multicolumn{6}{|c|}{DeepJIT with OPENSTACK}\\
        \hline
        \rowcolor[HTML]{EFEFEF} 
        $Test_{Avg}$ (OG) & \textbf{35} & \textbf{35} & \textbf{66} & \textbf{58} & \textbf{24} \\
        {\scriptsize \textit{Val. Min-Max }} & \scriptsize \textit{31-40} & \scriptsize \textit{31-40} & \scriptsize \textit{46-88} & \scriptsize \textit{47-63} & \scriptsize \textit{21-27} \\
        \hline
        \hline
        \rowcolor[HTML]{EFEFEF} 
        $Test_{Avg}$ (Platt) & \textbf{2} $\downdownarrows$ & \textbf{3}$\downdownarrows$ & \textbf{24} $\downdownarrows$ & \textbf{11} $\downdownarrows$ & \textbf{10} $\downdownarrows$ \\ 
        \hline
        {\scriptsize \textit{Val. Min-Max}} & \scriptsize \textit{1-4} & \scriptsize \textit{2-5} & \scriptsize \textit{3-59} & \scriptsize \textit{4-18} & \scriptsize \textit{8-11} \\
        \rowcolor[HTML]{f0e7e6}
        \hline
        {\scriptsize \textit{Stat. Sign. (OG-Platt) }} & \scriptsize \textit{Yes} & \scriptsize \textit{Yes} & \scriptsize \textit{Yes} & \scriptsize \textit{Yes} & \scriptsize \textit{Yes} \\
        \hline
        \hline
        \rowcolor[HTML]{EFEFEF} 
        $Test_{Avg}$ (Temp) & \textbf{36} $\up$ & \textbf{36} $\up$ & \textbf{60} $\down$ & \textbf{55} $\down$ & \textbf{24} $\no$ \\   
        \hline
        {\scriptsize \textit{Val. Min-Max}} & \scriptsize \textit{32-40} & \scriptsize \textit{32-40} & \scriptsize \textit{44-87} & \scriptsize \textit{45-61} & \scriptsize \textit{22-27} \\
        \hline
        \rowcolor[HTML]{f0e7e6}
        {\scriptsize \textit{Stat. Sign. (OG-Temp) }} & \scriptsize \textit{Yes} & \scriptsize \textit{Yes} & \scriptsize \textit{Yes} & \scriptsize \textit{Yes} & \scriptsize \textit{Yes} \\
        \hline        
        \hline
        \multicolumn{6}{|c|}{DeepJIT with QT}\\
        \hline
        \rowcolor[HTML]{EFEFEF} 
        $Test_{Avg}$ (OG) & \textbf{33} & \textbf{33} & \textbf{67} & \textbf{62} & \textbf{19} \\
        {\scriptsize \textit{Val. Min-Max }} & \scriptsize \textit{29-37} & \scriptsize \textit{29-37} & \scriptsize \textit{53-94} & \scriptsize \textit{51-64} & \scriptsize \textit{16-22} \\
        \hline
        \hline
        \rowcolor[HTML]{EFEFEF} 
        $Test_{Avg}$ (Platt) & \textbf{2} $\downdownarrows$ & \textbf{2} $\downdownarrows$ & \textbf{30} $\downdownarrows$ & \textbf{9} $\downdownarrows$ & \textbf{6} $\downdownarrows$ \\ 
        {\scriptsize \textit{Val. Min-Max}} & \scriptsize \textit{1-2} & \scriptsize \textit{1-3} & \scriptsize \textit{7-62} & \scriptsize \textit{2-14} & \scriptsize \textit{5-9} \\
        \hline
        \rowcolor[HTML]{f0e7e6}
        {\scriptsize \textit{Stat. Sign. (OG-Platt) }} & \scriptsize \textit{Yes} & \scriptsize \textit{Yes} & \scriptsize \textit{Yes} & \scriptsize \textit{Yes} & \scriptsize \textit{Yes} \\
        \hline
        \hline
        \rowcolor[HTML]{EFEFEF} 
        $Test_{Avg}$ (Temp) & \textbf{34} $\up$ & \textbf{34} $\up$ & \textbf{62} $\down$ & \textbf{59} $\down$ & \textbf{20} $\up$ \\ 
        \hline
        {\scriptsize \textit{Val. Min-Max}} & \scriptsize \textit{30-38} & \scriptsize \textit{30-38} & \scriptsize \textit{51-94} & \scriptsize \textit{49-60} & \scriptsize \textit{17-21} \\
        \hline
        \rowcolor[HTML]{f0e7e6}
        {\scriptsize \textit{Stat. Sign. (OG-Temp) }} & \scriptsize \textit{Yes} & \scriptsize \textit{Yes} & \scriptsize \textit{Yes} & \scriptsize \textit{Yes} & \scriptsize \textit{Yes} \\
        \hline 
        \hline
    \end{tabular}
    \label{combined_deepJIT}
\end{table}

\begin{figure}[htb]
\begin{tikzpicture}
  \node (img)  {\includegraphics[width=0.95\linewidth]{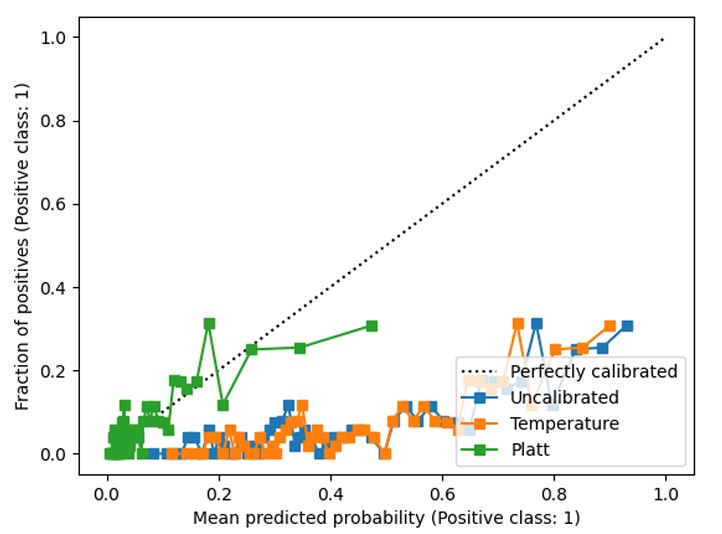}};
  \node[below=of img, node distance=0cm, yshift=1.25cm,font=\color{darkgrey}] {\footnotesize Confidence};
  \node[left=of img, node distance=0cm, rotate=90, anchor=center,yshift=-1cm,font=\color{darkgrey}] {\footnotesize Accuracy};
 \end{tikzpicture}
 \caption{Reliability Diagram (B=50, Adaptive) for DeepJIT  with QT dataset.}
 \label{fig:rel_diag_DeepJIT}
 \vspace*{-10pt}
\end{figure}

\hide{
\begin{figure}[htbp]
\centerline{\includegraphics[width=\linewidth]{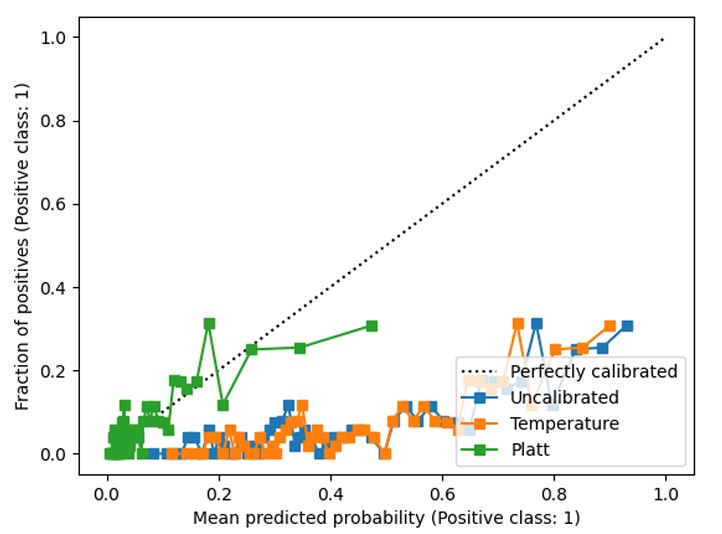}}
\caption{Reliability Diagram (B=50, adaptive) for DeepJIT  with QT dataset.}
\label{fig:rel_diag_DeepJIT}
\end{figure}}
Tables \ref{combined_deepJIT}, \ref{combined_LApredict}, and \ref{combined_CodeBERT} show the Expected Calibration Error (ECE), Maximum Calibration Error (MCE), and Brier scores for each JIT Defect Prediction technique. The reported values include measurements from the validation and test sets, both before and after calibration. 
The tables are organized as follows: reading the table top-down,  the first half presents the measurements for the OPENSTACK dataset (denoted as OS in the following text), while the second half presents those for the QT dataset.
Within each half, the calibration measurements are organized in the following order:
The first two rows contain the measurements for the original, uncalibrated JIT DP technique (denoted as OG).
The next two rows present the measurements after applying Platt scaling calibration (denoted as Platt). These are followed by the results of statistical tests assessing the significance of the differences between the original and Platt-calibrated measurements (denoted as Statistic. Sign).
The next rows contain the measurements after applying Temperature scaling calibration (denoted as Temp). Similarly, these are followed by the statistical test results evaluating the significance of the differences between the original and Temperature-calibrated measurements.\\
The first column indicates the dataset subset on which these measurements were computed.
The "$Test_{Avg}$" row reports for each calibration metric, the averaged measurements across all repetitions on the test sets. 
The "Val. Min-Max" row reports the minimum and maximum values across all validation folds to illustrate the measurements variance. \\
The subsequent columns in the table present the ECE and MCE values, using both \emph{equiwidth} and \emph{adaptive} binning schemas, and Brier scores in terms of rounded percentages. Due to space limitations, we only report results for B=15 bins; however, key findings related to B=50 are discussed below. 
The detailed tables are available in our repository \cite{b38}.

\subsubsection{DeepJIT}

Table \ref{combined_deepJIT}\footnote{The symbol "$\down$" indicates a miscalibration reduction by max 10\%.The symbol "$\downdownarrows$" indicates a miscalibration reduction by more than 10\%.The symbol "$\up$" indicates a miscalibration increase by max 10\%. The symbol "$\no$" indicates no change in calibration level.}  presents the calibration metrics for the DeepJIT model.
The ECE scores denote miscalibration rates of 35\% in the OS test set and 33\% in the QT test set. These values closely align with those observed in the validation sets.
Furthermore, the minimum and maximum values indicate low variance across cross-validation measurements.
We observed that the ECE scores are almost identical across different binning schemes (equiwidth vs. adaptive) and number of bins (15 vs. 50).
The consistency of the ECE scores across different binning configurations suggests that these scores accurately reflect DeepJIT's miscalibration levels.\\
DeepJIT exhibits an MCE score of 66\% (equiwidth) and 58\% (adaptive) with OS and 67\% and 62\% with QT test sets.\\ 
%We noticed some variability in MCE score measurements across test and validation sets.
In terms of Brier score, DeepJIT exhibits miscalibration rates of 24\% and 19\% in the OS and QT test sets, respectively. The measurements in the validation sets are comparable to these scores, with a variability of up to 6\%.\\
Analysis of the reliability diagrams from both validation and test sets indicates a stable pattern across cross-validation iterations.
The reliability diagram (the blue curve in Figure \ref{fig:rel_diag_DeepJIT}) reveals that DeepJIT generates predictions with probability estimates that are well distributed throughout the probability interval [0,1]. However, for most bins, the bin's accuracy is generally lower than its confidence, indicating DeepJIT's tendency to produce overconfident predictions probabilities.

\subsubsection{LApredict}
\begin{table}[h!]
    \caption{LApredict Calibration Scores}
    \centering    
    \renewcommand{\arraystretch}{1.2}
    \setlength{\arrayrulewidth}{0.1mm}
    \setlength{\tabcolsep}{3pt}
    \begin{tabular}{|c|c|c|c|c|c|}
        \hline
        \rowcolor[HTML]{EFEFEF} 
        \textbf{Data subset} & \multicolumn{2}{c|}{\textbf{ECE \%}} & \multicolumn{2}{c|}{\textbf{MCE \%}} & \textbf{Brier \%} \\
        & Equiwidth & Adaptive & Equiwidth & Adaptive & \\
        \rowcolor[HTML]{EFEFEF} 
        & 15 bins & 15 bins & 15 bins & 15 bins &  \\
        \hline
        \multicolumn{6}{|c|}{LApredict with OPENSTACK}\\
        \hline
        \rowcolor[HTML]{EFEFEF} 
        $Test_{Avg}$ (OG) & \textbf{9} & \textbf{8} & \textbf{21} & \textbf{12} & \textbf{15} \\
        %{Val. Avg (OG)} & 3 & 3 & 46 & 9 & 17 \\ 
        {\scriptsize \textit{Val. Min-Max }} & \scriptsize \textit{1-9} & \scriptsize \textit{2-8} & \scriptsize \textit{9-100} & \scriptsize \textit{4-17} & \scriptsize \textit{14-18} \\
        \hline    
        \hline
        \rowcolor[HTML]{EFEFEF} 
        $Test_{Avg}$ (Platt) & \textbf{8} $\down$ & \textbf{8} $\no$ & \textbf{22} $\up$ & \textbf{13} $\up$ & \textbf{15} $\no$ \\ 
        %{\scriptsize \textit{Test. Min-Max}} & \scriptsize \textit{7-10} & \scriptsize \textit{6-10} & \scriptsize \textit{19-25} & \scriptsize \textit{10-16} & \scriptsize \textit{14-15} \\
        \hline
        %{Val. Avg (Platt)} & 3 & 4 & 48 & 10 & 17 \\ 
        {\scriptsize \textit{Val. Min-Max}} & \scriptsize \textit{2-10} & \scriptsize \textit{2-10} & \scriptsize \textit{14-100} & \scriptsize \textit{4-18} & \scriptsize \textit{14-19} \\
        \hline
        \rowcolor[HTML]{f0e7e6}
        {\scriptsize \textit{Stat. Sign. (OG-Platt) }} & \scriptsize \textit{No} & \scriptsize \textit{No} & \scriptsize \textit{Yes} & \scriptsize \textit{No} & \scriptsize \textit{No} \\
        \hline
        \hline
        \rowcolor[HTML]{EFEFEF} 
        $Test_{Avg}$ (Temp) & \textbf{9} $\no$ & \textbf{9} $\up$ & \textbf{21} $\no$ & \textbf{13} $\up$ & \textbf{15} $\no$ \\   
        %{\scriptsize \textit{Test. Min-Max}} & \scriptsize \textit{7-11} & \scriptsize \textit{7-11} & \scriptsize \textit{19-23} & \scriptsize \textit{12-15} & \scriptsize \textit{14-15} \\
        \hline
        %{Val. Avg (Temp)} & 3 & 3 & 47 & 9 & 17 \\ 
        {\scriptsize \textit{Val. Min-Max}} & \scriptsize \textit{2-11} & \scriptsize \textit{2-11} & \scriptsize \textit{12-100} & \scriptsize \textit{4-15} & \scriptsize \textit{14-19} \\
        \hline
        \rowcolor[HTML]{f0e7e6}
        {\scriptsize \textit{Stat. Sign. (OG-Temp) }} & \scriptsize \textit{No} & \scriptsize \textit{No} & \scriptsize \textit{No} & \scriptsize \textit{Yes} & \scriptsize \textit{No} \\
        \hline
        \hline
        \multicolumn{6}{|c|}{LApredict with QT}\\
        \hline
        \rowcolor[HTML]{EFEFEF} 
        $Test_{Avg}$ (OG) & \textbf{3} & \textbf{2} & \textbf{99} & \textbf{8} & \textbf{12} \\
        %{Val. Avg (OG)} & 2 & 3 & 70 & 8 & 12 \\ 
        {\scriptsize \textit{Val. Min-Max }} & \scriptsize \textit{1-4} & \scriptsize \textit{2-4} & \scriptsize \textit{21-100} & \scriptsize \textit{3-18} & \scriptsize \textit{11-13} \\
        \hline      
        \hline
        \rowcolor[HTML]{EFEFEF} 
        $Test_{Avg}$ (Platt) & \textbf{3} $\no$ & \textbf{2} $\no$ & \textbf{88} $\downdownarrows$ & \textbf{8} $\no$ & \textbf{12} $\no$ \\ 
        %{\scriptsize \textit{Test. Min-Max}} & \scriptsize \textit{1-2} & \scriptsize \textit{2-3} & \scriptsize \textit{47-100} & \scriptsize \textit{4-13} & \scriptsize \textit{12-12} \\
        %{Val. Avg (Platt)} & 2 & 3 & 66 & 9 & 12 \\ 
        {\scriptsize \textit{Val. Min-Max}} & \scriptsize \textit{1-5} & \scriptsize \textit{2-5} & \scriptsize \textit{16-100} & \scriptsize \textit{3-22} & \scriptsize \textit{11-13} \\
        \hline
        \rowcolor[HTML]{f0e7e6}
        {\scriptsize \textit{Stat. Sign. (OG-Platt) }} & \scriptsize \textit{No} & \scriptsize \textit{No} & \scriptsize \textit{Yes} & \scriptsize \textit{No} & \scriptsize \textit{No} \\
        \hline        
        \hline
        \rowcolor[HTML]{EFEFEF} 
        $Test_{Avg}$ (Temp) & \textbf{2} $\down$ & \textbf{2} $\no$ & \textbf{99} $\no$ & \textbf{9} $\up$ & \textbf{12} $\no$ \\ 
        %{\scriptsize \textit{Test. Min-Max}} & \scriptsize \textit{1-3} & \scriptsize \textit{2-3} & \scriptsize \textit{98-100} & \scriptsize \textit{7-11} & \scriptsize \textit{12-12} \\
        \hline
        %{Val. Avg (Temp)} & 2 & 3 & 69 & 8 & 12 \\ 
        {\scriptsize \textit{Val. Min-Max}} & \scriptsize \textit{1-4} & \scriptsize \textit{2-4} & \scriptsize \textit{17-100} & \scriptsize \textit{4-15} & \scriptsize \textit{11-13} \\
        \hline
        \rowcolor[HTML]{f0e7e6}
        {\scriptsize \textit{Stat. Sign. (OG-Temp) }} & \scriptsize \textit{No} & \scriptsize \textit{No} & \scriptsize \textit{No} & \scriptsize \textit{Yes} & \scriptsize \textit{No} \\
        \hline
        
        \hline
    \end{tabular}
    \label{combined_LApredict}
\end{table}

Table \ref{combined_LApredict} presents the calibration metrics for the LApredict model.
The ECE scores denote miscalibration rates of 9\% (equiwidth) and 8\% (adaptive) on the OS test set, and  3\% and 2\% on the QT test set. 
The ECE measurements during cross-validation show a min-max variance of up to 8\%. \\
In terms of the MCE score LApredict model exhibits miscalibration rates of up to 21\% (equiwidth) and 12\% (adaptive) on the OS test set, and 99\% and 8\% on the QT test set. 
Analysis of the MCE across different binning schemes indicates that the adaptive binning schema generally yields lower scores, denoting better calibration levels, than the equiwidth schema.
It is evident that the calibration performance of LApredict varies considerably between the OS and QT datasets. Specifically, when trained on the OS dataset, LApredict exhibits greater miscalibration in terms of ECE score, whereas on the QT dataset results in higher miscalibration as measured by the MCE score.
To investigate this behavior, we examined the bin sizes and their distributions. Our analysis revealed that LApredict produces fewer empty bins on the OpenStack dataset compared to the QT dataset ($~3x$ empty bins), indicating a better distribution of predictions across bins. 
\begin{figure}[htb] 
{\includegraphics[width=0.95\linewidth]{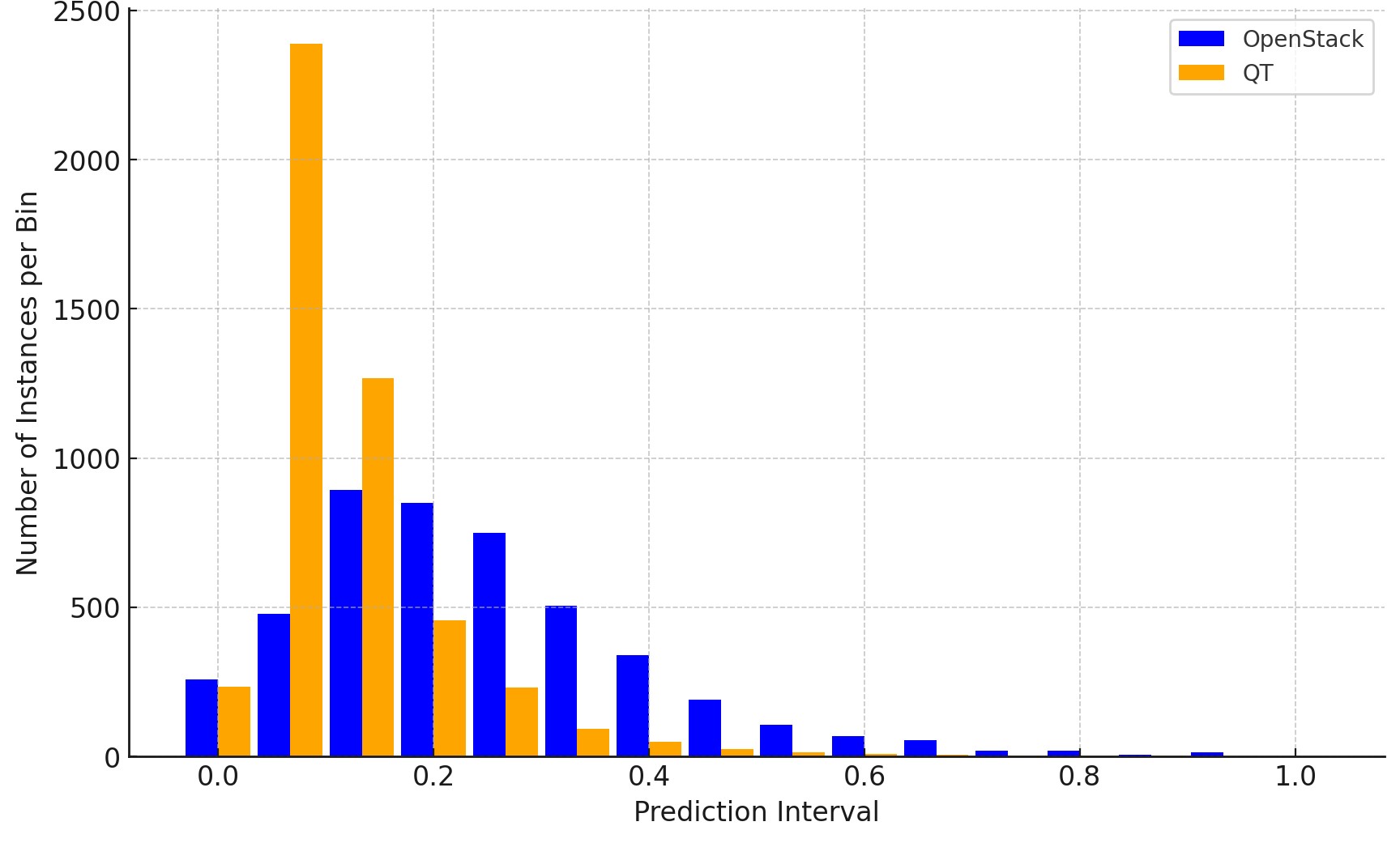}}
 \caption{Bin sizes for LApredict with (B=15, Equiwidth) for LApredict with both datasets.}
 \label{fig:LApredict_bins} 
\vspace*{-10pt}
\end{figure}
However, as illustrated in Figure \ref{fig:LApredict_bins}, in both cases, predictions tend to cluster in the leftmost bins—those associated with the lowest confidence levels [0,0.5]. This skewness is more pronounced for the QT dataset than for OS, contributing to its higher MCE scores. Specifically, for the OS dataset, we observed that nearly 70\% of predictions fall within the bins spanning across the first half of the confidence interval, whereas for the QT dataset, this proportion increases up to 96\%.
This difference suggests that the model generalizes differently across these datasets, and LApredict with QT has a higher tendency of predicting instances as \emph{\rmfamily{clean}}.
Examining the model's accuracy metrics, we noticed that, while the AUC scores are similar for both datasets (0.75 for OS and 0.74 for QT), there is a clear difference in other performance metrics. Specifically, LApredict achieves a precision of 0.68 and a recall of 0.078 for the OS dataset, versus 0.37 precision and 0.024 recall for the QT dataset.\\
The reliability diagram (the blue curve in Figure \ref{fig:rel_diag_LApredict}) further illustrates the miscalibration of LApredict.
The curve indicates that LApredict is well-calibrated in the left-most bins, where the majority of predictions are concentrated, but not consistently across the entire prediction confidence interval. For confidence scores above 0.3, LApredict exhibits both overconfidence in certain probability ranges and underconfidence in others.
Although LApredict achieves a notably lower ECE score compared to other JIT DP techniques, the reliability diagram highlights pronounced fluctuations. These alternating overconfident and underconfident regions balance out, leading to relatively low ECE scores. However, the high MCE value reflects the severity of these fluctuations.

\begin{figure}[htb] 
\begin{tikzpicture}
  \node (img)  {\includegraphics[width=0.95\linewidth]{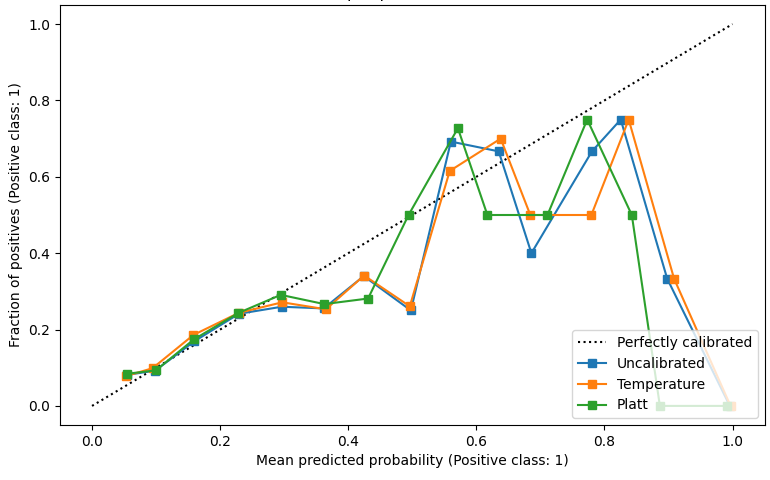}};
  \node[below=of img, node distance=0cm, yshift=1.25cm,font=\color{darkgrey}] {\footnotesize Confidence};
  \node[left=of img, node distance=0cm, rotate=90, anchor=center,yshift=-1cm,font=\color{darkgrey}] {\footnotesize Accuracy};
 \end{tikzpicture}
 \caption{Reliability Diagram (B=15, Equiwidth) for LApredict with QT dataset.}
 \label{fig:rel_diag_LApredict} 
\vspace*{-10pt}
\end{figure}

\subsubsection{CodeBERT4JIT}

\hide{
\begin{figure}[htbp]
\centerline{\includegraphics[width=\linewidth]{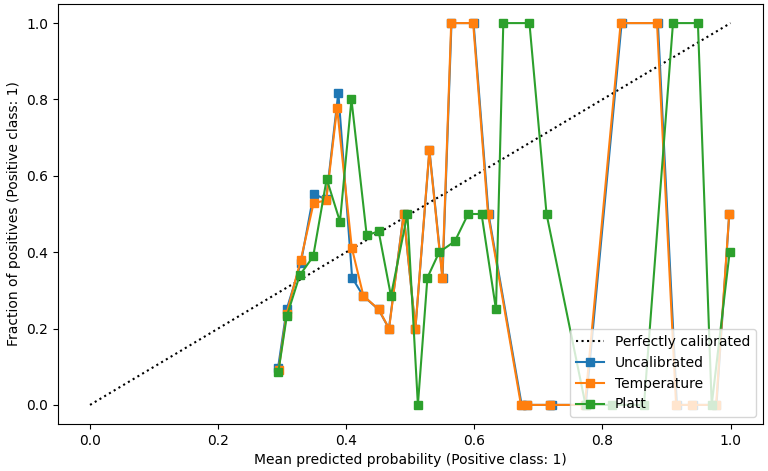}}
\caption{Reliability Diagram (B=50, equiwidth) for LApredict with QT dataset. }
\label{fig:rel_diag_LApredict} 
\end{figure}}

Table \ref{combined_CodeBERT} shows the measured calibration metrics for CodeBERT4JIT model.
The ECE scores denote miscalibration rates of 12\% (equiwidth) and 11\% (adaptive) in the OS test set and 8\% and 7\% in the QT test set. 
ECE measurements during cross-validation show a (min-max) variance of up to 7\%.\\
In terms of MCE scores, CodeBERT4JIT exhibits miscalibration rates of 70\% (equiwidth) and 43\% (adaptive) in the OS test set and 74\% and 30\% in the QT test set.
By observing the MCE scores across different binning schemas we notice that the adaptive binning schema generally yields lower scores than the equiwidth schema.\\
In terms of Brier score, CodeBERT4JIT exhibits 12\% and 8\% miscalibration rate with OS and QT test sets, respectively. \\
The reliability diagram (the blue curve in Figure \ref{fig:rel_diag_CodeBERT}) illustrates that the model is underconfident when prediction probabilities fall in the range [0 - 0.1] but overconfident across the rest of the probability spectrum.
A closer examination of the contents of bins from various data subsets and binning configurations reveals a consistent trend across iterations: the outermost bins contain the majority of predictions, while the remaining bins hold only a few, or sometimes no, predictions. 
We observe that the first bin, the leftmost one, which includes instances the model predicts as \emph{\rmfamily{clean}} with high probability, contains nearly all predictions. 
This occurrence explains why CodeBERT4JIT exhibits underconfidence in that specific region. Moreover, this also suggests that CodeBERT4JIT is biased toward classifying commits as \emph{\rmfamily{clean}}.

\subsubsection*{General observations}
We observed, in all our experiments, that when setting the number of bins to 50 for computing the calibration metrics, the JIT defect prediction models exhibit higher miscalibration rates.
A similar trend is also observed for the MCE score with equiwidth vs. adaptive binning schema, where the equiwidth schema typically yields higher miscalibration rates.  

\begin{tcolorbox}[sharp corners, colback=lightgray!5, colframe=gray!90!blue!90!, title= \ding{101} Answering Research Question 1:]
Our experimental results indicate that all JIT DP models under investigation exhibit some degree of miscalibration, with Expected Calibration Error ranging from 2\% to 35\%, and Brier score ranging from 8\% to 24\%. Our experiments indicate that the degree of miscalibration varies across different datasets.
\end{tcolorbox}

\subsection{RQ2: To what extent can post-calibration methods improve the calibration levels of existing JIT DP models?}

\subsubsection{DeepJIT}
\begin{table}[h!]
    \caption{CodeBERT4JIT Calibration Scores}
    \centering    
    \renewcommand{\arraystretch}{1.2}
    \setlength{\arrayrulewidth}{0.1mm}
    \setlength{\tabcolsep}{3pt}
    \begin{tabular}{|c|c|c|c|c|c|}
        \hline
        \rowcolor[HTML]{EFEFEF} 
        \textbf{Data subset} & \multicolumn{2}{c|}{\textbf{ECE \%}} & \multicolumn{2}{c|}{\textbf{MCE \%}} & \textbf{Brier \%} \\
        & Equiwidth & Adaptive & Equiwidth & Adaptive & \\
        \rowcolor[HTML]{EFEFEF} 
        & 15 bins & 15 bins & 15 bins & 15 bins &  \\
        \hline
        \multicolumn{6}{|c|}{CodeBERT4JIT with OPENSTACK}\\
        \hline
        \rowcolor[HTML]{EFEFEF} 
        T$Test_{Avg}$ (OG) & \textbf{12} & \textbf{11} & \textbf{70} & \textbf{43} & \textbf{12} \\
        {\scriptsize \textit{Validation Min-Max }} & \scriptsize \textit{8-15} & \scriptsize \textit{8-15} & \scriptsize \textit{50-92} & \scriptsize \textit{22-65} & \scriptsize \textit{10-16} \\
        \hline
        \hline
        \rowcolor[HTML]{EFEFEF} 
        $Test_{Avg}$ (Platt) & \textbf{4} $\down$ & \textbf{4} $\down$ & \textbf{49} $\downdownarrows$ & \textbf{14} $\downdownarrows$ & \textbf{10} $\down$ \\ 
        \hline
        {\scriptsize \textit{Validation Min-Max}} & \scriptsize \textit{1-6} & \scriptsize \textit{2-6} & \scriptsize \textit{11-90} & \scriptsize \textit{6-25} & \scriptsize \textit{8-11} \\
        \hline
        \rowcolor[HTML]{f0e7e6}
        {\scriptsize \textit{Stat. Sign. (OG-Platt) }} & \scriptsize \textit{Yes} & \scriptsize \textit{Yes} & \scriptsize \textit{Yes} & \scriptsize \textit{Yes} & \scriptsize \textit{Yes} \\
        \hline
        \hline
        \rowcolor[HTML]{EFEFEF} 
        $Test_{Avg}$ (Temp) & \textbf{6} $\down$ & \textbf{6} $\down$ & \textbf{64} $\down$ & \textbf{30} $\downdownarrows$ & \textbf{11} $\down$ \\   
        \hline
        {\scriptsize \textit{Validation Min-Max}} & \scriptsize \textit{2-14} & \scriptsize \textit{3-14} & \scriptsize \textit{37-95} & \scriptsize \textit{8-50} & \scriptsize \textit{9-14} \\
        \hline
        \rowcolor[HTML]{f0e7e6}
        {\scriptsize \textit{Stat. Sign. (OG-Temp) }} & \scriptsize \textit{Yes} & \scriptsize \textit{Yes} & \scriptsize \textit{Yes} & \scriptsize \textit{Yes} & \scriptsize \textit{Yes} \\
        \hline       
        \hline
        \multicolumn{6}{|c|}{CodeBERT4JIT with QT}\\
        \hline
        \rowcolor[HTML]{EFEFEF} 
        $Test_{Avg}$ (OG) & \textbf{8} & \textbf{7} & \textbf{74} & \textbf{30} & \textbf{8} \\
        {\scriptsize \textit{Validation Min-Max }} & \scriptsize \textit{6-11} & \scriptsize \textit{5-11} & \scriptsize \textit{50-91} & \scriptsize \textit{13-73} & \scriptsize \textit{6-11} \\
        \hline
        \hline
        \rowcolor[HTML]{EFEFEF} 
        $Test_{Avg}$ (Platt) & \textbf{2} $\down$ & \textbf{3} $\down$ & \textbf{53} $\downdownarrows$ & \textbf{9} $\downdownarrows$ & \textbf{6} $\down$\\ 
        {\scriptsize \textit{Validation Min-Max}} & \scriptsize \textit{1-3} & \scriptsize \textit{1-4} & \scriptsize \textit{12-95} & \scriptsize \textit{4-20} & \scriptsize \textit{5-7} \\
        \hline
        \rowcolor[HTML]{f0e7e6}
        {\scriptsize \textit{Stat. Sign. (OG-Platt) }} & \scriptsize \textit{Yes} & \scriptsize \textit{Yes} & \scriptsize \textit{Yes} & \scriptsize \textit{Yes} & \scriptsize \textit{Yes} \\
        \hline
        \hline
        \rowcolor[HTML]{EFEFEF} 
        $Test_{Avg}$ (Temp) & \textbf{4} $\down$ & \textbf{4} $\down$ & \textbf{67} $\down$ & \textbf{26} $\down$ & \textbf{7} $\down$ \\ 
        \hline
        {\scriptsize \textit{Validation Min-Max}} & \scriptsize \textit{2-10} & \scriptsize \textit{2-10} & \scriptsize \textit{42-97} & \scriptsize \textit{3-55} & \scriptsize \textit{5-10} \\
        \hline
        \rowcolor[HTML]{f0e7e6}
        {\scriptsize \textit{Stat. Sign. (OG-Temp) }} & \scriptsize \textit{Yes} & \scriptsize \textit{Yes} & \scriptsize \textit{Yes} & \scriptsize \textit{Yes} & \scriptsize \textit{Yes} \\
        \hline     
        \hline
    \end{tabular}
    \label{combined_CodeBERT}
\end{table}

Table \ref{combined_deepJIT} presents the calibration metrics for the DeepJIT model after applying Platt scaling and Temperature scaling.

\emph{Platt scaling}: reduces the ECE miscalibration of DeepJIT on  the OS test set from 35\% (original DeepJIT across both binning schemes) to 2\% (equiwidth) and 3\% (adaptive). Similarly, for the QT test set, the ECE scores drop from 33\% to 2\%, for both binning schemes.\\
Similarly, after applying Platt scaling, we observe lower MCE scores for both datasets across all binning configurations. \\
The DeepJIT's miscalibration in terms of Brier score also decreases after applying Platt scaling, dropping from 24\% to 10\% for OS and from 19\% to 6\% for QT test sets.\\
The computed statistical tests, suggest that the effect of Platt scaling on DeepJIT's miscalibration is statistically significant across all the measured metrics.\\
Reliability diagrams after applying Platt scaling (the green curve in Figure \ref{fig:rel_diag_DeepJIT}) reveal that the reliability curve aligns more closely with the diagonal, representing ideal calibration. 
However, we notice that Platt scaling compresses the predicted probabilities for the C=1 class, i.e., the commits classified as \emph{\rmfamily{defective}}, to the first half of the x-axis (below 0.5), indicating an adjustment of the probabilities distribution for this class.
\begin{figure}[htb]
\begin{tikzpicture}
  \node (img)  {\includegraphics[width=0.95\linewidth]{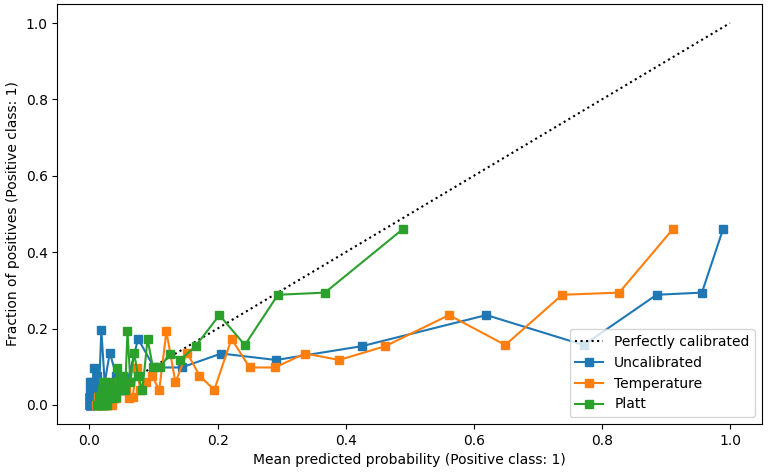}};
  \node[below=of img, node distance=0cm, yshift=1.25cm,font=\color{darkgrey}] {\footnotesize Confidence};
  \node[left=of img, node distance=0cm, rotate=90, anchor=center,yshift=-1cm,font=\color{darkgrey}] {\footnotesize Accuracy};
 \end{tikzpicture}
 \caption{Reliability Diagram (B=50, Adaptive) for CodeBERT4JIT with QT dataset.}
 \label{fig:rel_diag_CodeBERT}
 \vspace*{-10pt}
\end{figure}

\emph{Temperature scaling}: does not substantially improve the calibration of DeepJIT model. 
After applying Temperature scaling, the ECE miscalibration scores exhibit a slight increase by 1\% for both OS and QT datasets.\\
However, Temperature scaling does produce lower MCE values compared to the uncalibrated DeepJIT.\\
The Brier scores remain almost consistent with prior results, at 24\% for OS and 20\% (+1\%) for QT.\\
However, the computed statistical tests, suggest that the effect of Temperature scaling on DeepJIT's miscalibration is statistically significant across all the measured metrics.\\
The reliability diagram (the orange curve in Figure \ref{fig:rel_diag_DeepJIT}) reveals that Temperature slightly compresses the plotted curve in the center of the x-axis; however, it does not cause a substantial deviation from the uncalibrated DeepJIT curve.

\subsubsection{LApredict}

Table \ref{combined_LApredict} presents the calibration metrics for the LApredict model after applying Platt scaling and Temperature scaling.

\emph{Platt scaling}: has a marginal effect on the miscalibration of LApredict, resulting in fluctuations of approximately ±1\% in the ECE score across both datasets.\\
In terms of the MCE scores the results for both datasets are nearly the same with those of the original LApredict model.\\
Similarly, Platt scaling does not notably affect the Brier score of LApredict. 
For both datasets, the Brier scores remain constant, indicating no change in miscalibration.\\
The computed statistical tests indicate that the effect of Platt scaling on LApredict's miscalibration is generally not statistically significant. Statistical significance is observed only for some of the configurations of the MCE measurements on the test set; however, these results are not consistent with the statistical significance test outcomes for the validation set measurements.\\
The reliability diagram in Figure \ref{fig:rel_diag_LApredict} indicates that Platt scaling, slightly shifts the original curve, with minor deviations. However, it does not correct the existing peaks and valleys.

\emph{Temperature scaling}: also induces a marginal effect on the miscalibration of LApredict, resulting in fluctuations of approximately ±1\% in the ECE score across both datasets.\\
Regarding MCE scores, we observe the same effect, where Temperature scaling typically leads to MCE scores that are nearly the same as those of the original LApredict model, for both datasets.\\
In terms of Brier scores, the Temperature-scaled LApredict exhibits miscalibration scores consistent with those of the uncalibrated model.\\
Same as for Platt scaling, the statistical tests indicate that the effect of Temperature scaling on LApredict's miscalibration is generally not statistically significant. Statistical significance is observed only for some configurations of the MCE measurements on the test set.\\
The reliability diagram in Figure \ref{fig:rel_diag_LApredict} supports the above-mentioned observations. As we can see the curve of Lapredict with Temperature scaling, shows only minor deviations form the curve of the uncalibrated model, indicating that Temperature scaling has minor effects on the calibration of LApredict.
\hide{
\begin{figure}[htbp]
\centerline{\includegraphics[width=\linewidth]{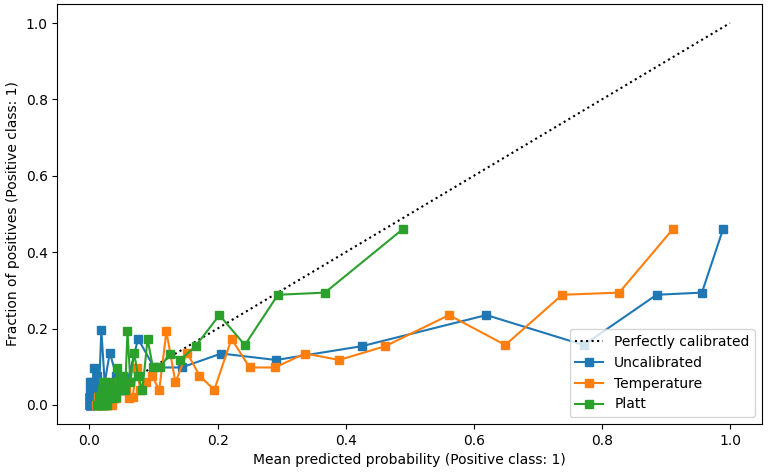}}
\caption{Reliability Diagram (B=50, adaptive) for CodeBERT4JIT with QT dataset.}
\label{fig:rel_diag_CodeBERT}
\end{figure}}

\subsubsection{CodeBERT4JIT}
Table \ref{combined_CodeBERT} presents the calibration metrics for the CodeBERT4JIT model after applying Platt scaling and Temperature scaling.

\emph{Platt scaling}: notably improves the miscalibration of the CodeBERT4JIT model. Specifically, Platt scaling reduces the ECE on the OS test set from 12\% (equiwidth) to 4\%, and from 11\% (adaptive), to 4\%. Similarly, for the QT dataset, the ECE miscalibration is reduced from 8\% to 2\% (equiwidth) and from 7\% to 3\% (adaptive), after applying Platt scaling.\\
The same effect can be observed also for the MCE scores on both datasets and across all binning schemas.\\
The CodeBERT4JIT's miscalibration in terms of Brier score also decreases after applying Platt scaling, droping from 12\% to 10\% for OS and from 8\% to 6\% for QT test sets.\\ 
The results of the statistical tests indicate that Platt scaling has a statistically significant impact on CodeBERT4JIT’s miscalibration across all evaluated metrics.\\
Reliability diagrams after applying Platt scaling (Figure \ref{fig:rel_diag_CodeBERT}) reveal that the reliability curve aligns more closely with the diagonal, indicating that the accuracy of the bins better reflects their confidence.
However, here again, we observe the same pattern as for DeepJIT, where applying Platt compresses the reliability curve to the left-most half of the x-axis.

\emph{Temperature scaling}: also causes a reduction on CodeBERT4JIT's miscalibration, however, not to the same degree as Platt scaling.
Specifically, applying Temperature scaling, leads to a reduction, in terms of the ECE score, from 12\% (equiwidth) to 6\%, and from 11\% (adaptive) to 6\% for OS. Similarly for QT we observe a reduction of the ECE scores from 8\% to 4\% and from and 7\% to 4\%.\\
Temperature scaling also leads to lower MCE scores across both datasets and binning schemes.\\ 
Temperature scaling slighlty decreases Brier scores, denoting a drop by 1\% for both OPENSTACK and QT datasets.\\
The results of the statistical tests indicate that Temperature scaling has a statistically significant impact on CodeBERT4JIT’s miscalibration across all evaluated metrics.\\
The reliability curve in Figure \ref{fig:rel_diag_CodeBERT} shows that Temperature scaling, slightly shits the plotted curve closer to the ideal calibration curve, compared to the uncalibrated CodeBERT4JIT curve. However, the curve still remains far from being well-calibrated.

\begin{tcolorbox}[sharp corners, colback=lightgray!5, colframe=gray!90!blue!90!, title=\ding{101} Answering Research Question 2:]

Platt scaling generally improved calibration by reducing miscalibration scores across different models, with the most significant improvements seen in DeepJIT and CodeBERT4JIT, with an observed ECE score reduction of up to 33\% and 8\% respectively. However, its impact on LApredict was minimal. Temperature scaling showed mixed effects - while it improved the miscalibration rates of CodeBERT4JIT, it showed varied results for DeepJIT and LApredict.  
The effects of both Platt and Temperature scaling on DeepJIT and CodeBERT4JIT were statistically significant across all measured metrics. However, for LApredict, their statistical significance was observed in only a few of the measured metrics.

\end{tcolorbox}

\section{Discussion and Implications}
Our experimental results indicate that all evaluated Just-in-Time defect prediction models exhibit some degree of miscalibration, across all measured calibration metrics. Among the evaluated models, DeepJIT demonstrated the highest level of miscalibration, while LApredict exhibited the lowest.
Prior research on model calibration suggests that miscalibration tends to be more pronounced in complex machine learning models \cite{b17,b18}. Consequently, given that LApredict is a simpler model based on logistic regression, it was expected to achieve lower miscalibration scores compared to more complex models. However, our findings did not consistently support this expectation. For instance, on the OPENSTACK dataset, LApredict’s ECE scores were comparable to those of CODEBERT4JIT with QT.  Given LApredict’s simplicity and the high class imbalance present in the OPENSTACK dataset, we hypothesize that the model struggles to generalize effectively.

Moreover, we observed that each of these models exhibits different calibration levels when trained on different datasets. Specifically, the OPENSTACK dataset generally results in higher miscalibration compared to QT dataset.
We attribute this to the smaller size of the OPENSTACK dataset, which promotes overfitting, particularly given the pronounced class imbalance.

Our experimental findings emphasize the importance of assessing and, if necessary, correcting model calibration prior to using prediction probabilities as confidence scores to guide quality assurance decision-makers, for example, in determining whether a prediction is correct or how to prioritize predictions.

Our experimental findings do not provide conclusive evidence that post-calibration methods improve the calibration of existing JIT defect prediction techniques. Although, in some cases, Platt scaling did significantly reduce miscalibration error (ECE, MCE, and Brier score), the reliability diagrams showed that it also tends to compress the predicted probabilities for class C=1 (\emph{\rmfamily{defective}}) into the first half of the probability interval. This means that the model typically assigns probabilities below 0.5 for the \emph{\rmfamily{defective}} class, resulting in a low likelihood of classifying any commits as \emph{\rmfamily{defective}}, if the classification threshold is set at 0.5.
To address this bias, careful optimization of the classification threshold would be needed.\\
However, for the simpler JIT defect prediction model (LApredict), Platt scaling did not effectively reduce model miscalibration. This outcome was expected, as LApredict is inherently a logistic regression model; thus, applying Platt scaling - another logistic model -was unlikely to yield significant improvements. Furthermore, LApredict already exhibited relatively low miscalibration scores.

Temperature scaling showed limited capabilities in improving JIT defect prediction model's calibration. Our experimental results indicated that Temperature scaling induces only minor deviations from the original, uncalibrated metrics.
A promising direction for future work would be to explore design-time calibration methods. This involves integrating the calibration process directly into model training, often by using specialized loss functions or regularization techniques that promote well-calibrated prediction probabilities \cite{b28}.

Our findings consistently show that increasing the number of bins when computing calibration metrics, results in higher miscalibration scores.
We also observed that the equiwidth binning schema typically produces higher MCE miscalibration scores compared to the adaptive binning schema.
This suggests that the predictions are not evenly distributed, but rather tend to bundle up in specific regions of the prediction probability interval.
However, this does not imply that a specific binning configuration -such as using fewer bins with adaptive binning- is inherently better than another configuration, like using more bins with equiwidth binning.
A small number of bins leads to more instances grouped within each bin, which can reduce ECE and MCE due to the stabilizing effect of averaging over larger sample sizes. 
In contrast, a higher number of bins results in fewer instances per bin, sometimes leaving bins empty. This sparse distribution highlights "edge" cases, leading to higher miscalibration scores and greater variability; especially with biased models, like CodeBERT4JIT, or with highly imbalanced datasets, like QT.
A similar effect is observed with adaptive binning schema, which equalizes the number of instances in each bin, thereby smoothing and reducing the observed miscalibration via the averaging effect.

Choosing a binning configuration should depend on the desired analysis goal. If a detailed view of predicted probabilities to detect clusters and edge cases is needed, a large number of bins with equiwidth binning may be preferable. Conversely, if a general assessment of model miscalibration is the focus, fewer bins may provide a more stable measure.

It is important to note that ECE, MCE, and reliability diagrams are designed for models that are expected to produce dispersed probability estimates, i.e., estimates that are well spread across the probability interval. However, this may not always be preferable. For instance, in Just-In-Time defect prediction, where predictions are binary, sharpness -rather than dispersion of probability estimates- may be preferred.
Sharpness refers to prediction probability estimates that are primarily concentrated near zero or one, with minimal occurrences of intermediate values \cite{b36}. Such a distribution can facilitate the identification of incorrect predictions. The Brier score is commonly used as a calibration metric to measure the proximity of predicted probabilities to the actual class of the input. However, achieving sharpness remains challenging due to imperfect JIT defect predictors.

\section{Validity risks}
\hide{Internal: 
In this study, we utilize three existing JIT defect prediction techniques, each employing distinct model architectures, to minimize potential selection bias. 
To mitigate replicability errors, we reuse the original implementations from the authors of the corresponding publications.
Following the replication guidelines from the respective GitHub repositories, we executed the code and confirmed that each JIT DP technique achieves the accuracy levels reported in the original publication. \todo{remove!}
\textcolor{blue}{Also, we follow open-source implementations as much as possible, to avoid any implementation errors and we used independent experts to ensure correctness of implementation and data ;-)}
\todo{LOOLLL, independent experts :'D} }

\emph{Internal:}
Certain calibration metrics used in this study, such as ECE, MCE, and reliability diagrams, are parametric; thus, sensitive to the choice of the number of bins, and the binning schema (see section \ref{calibration_metrics}).
%\textcolor{blue}{Remove the detailed explanation of parametric sensitivity}
% Remove from here
Using a high number of bins can provide a very granular view, capturing fine details of the model's calibration at different probability ranges.
%However, if bins are too small, some bins may contain few, or no predictions. This causes a high variance of the estimated accuracy within those bins, potentially distorting the overall calibration score. 
A small number of bins leads to an increasing number of predictions per bin; thereby, reducing the variance within each bin, yielding a smoother, more stable reliability diagram, though it may mask miscalibrations in specific probability regions. 
% until here
To address this risk, we experiment with several bin sizes. We draw these values from existing literature.
Additionally, we experiment with different binning schemas to assess if the JIT DP models generate predictions with evenly distributed, balanced probabilities across different bins or if they show a tendency to cluster within certain probability ranges. 
Moreover, we employ the Brier score, a non-parametric, bin-independent metric, to further assess the models' calibration.

Defect datasets are typically characterized by class imbalances, increasing the risk that the validation set may lack instances belonging to the minority class. Such a scenario can lead to models that falsely exhibit  good calibration, even if they have overfitted the majority class. 
To mitigate this risk, we performed cross-validation, using each fold of the training data as a validation set at least once. Additionally, we evaluated model calibration on the test set to ensure that we measure the calibration across all subsets of the subject datasets.

\emph{External:} 
This is limited to the investigation of only three state-of-the-art JIT defect prediction techniques, each evaluated with two datasets. This constrains the generalizability of our findings to other JIT defect prediction techniques and datasets. Expanding the scope of techniques and datasets in future studies may yield more comprehensive insights into the generalizability of these results.

\section{Related work}

%Just-in-Time Defect Prediction remains an active research domain, with numerous studies published annually.aiming at improving both the accuracy and practical utility of these tools. 
Several JIT DP approaches have proposed leveraging prediction probabilities to refine predictions, with an emphasis on ranking and prioritizing defect-prone commits. However, these probability-based ranking methods presuppose that predictive models are well-calibrated, which typically remains unverified.

Arisholm and Briand~\cite{b10,b14} propose prioritizing code segments predicted as defective code by ranking them based on the respective prediction probabilities.
%favoring code fragments with fewer lines (presumed to incur lower investigation costs). 
Effort-aware defect prediction extends this approach by ranking predictions not solely on probability but on investigation effort; typically calculated as a ratio between the probability that a commit is defective and the commit complexity; thereby, prioritizing highly-likely defective, low-effort commits~\cite{b13,b15,b16}.

Mequita et al.~\cite{b9} propose utilizing prediction probabilities as confidence scores. They use these scores to identify predictions made with low confidence and flag them as potential misclassifications. They build a defect prediction model with a rejection mechanism that calculates the prediction probability for each prediction. If a prediction's probability is below a predefined confidence threshold, the model rejects it and refers the decision to a practitioner. However, the authors do not evaluate the calibration of the model before relying on prediction probabilities for rejecting predictions.

Shahini et al.~\cite{b6} recognize that relying on prediction probabilities to differentiate correct from incorrect predictions can be misleading. The authors proposed using conformal prediction (CP) to identify JIT defect predictions that can be guaranteed to be correct. The CP algorithm takes as input prediction probabilities to compute rigorous measures of prediction confidence that can accurately detect correct predictions. However, the approach does not take into account that the JIT defect orediction models might be badly calibrated, before feeding the prediction probabilities into the CP algorithm.

One study that explicitly acknowledges the risk of miscalibrated probabilities in defect prediction is presented by Wan et al. \cite{b6}.
They introduce a metric called the Adjusted Trust Score (ATS) to evaluate the likelihood that each prediction may be a misclassification. The ATS score combines the predicted probability of the class with the input's distance from the training data.
Using the ATS score, they build a defect prediction model with rejection. The model rejects all predictions with an ATS below a specified threshold, marking them as potential misclassifications. 
The authors employ Platt scaling to improve the calibration of the defect prediction model.  However, they do not evaluate the model's calibration either before or after implementing Platt scaling.

\section{CONCLUSION}
%$\ding{101}$
\hide{
Just-in-Time Defect Prediction remains an active research domain, with numerous studies published annually aiming at improving both the accuracy and practical utility of these tools. 
Despite recent advances, which have contributed significantly in improving the accuracy of JIT DP, these tools still produce a significant rate of false positive or false
negative predictions. To overcome this challenge, one can extract alongside each prediction, the corresponding prediction probability. These probabilities can enable the detection of JIT DP misclassifications, and prioritize predictions for better QA resources allocation. However, for these probabilities to be reliable, the model needs to be well calibrates.}

This study presents an empirical analysis of the calibration accuracy of three state-of-the-art Just-in-Time defect prediction techniques. The study's goal was to assess whether the prediction probabilities generated by JIT defect prediction models can be used to reliably guide quality assurance practitioners on whether to trust JIT defect predictions, and how to efficiently prioritize them. 
We utilized several established calibration metrics to measure the extent of miscalibration of the selected JIT defect prediction models. 

Our experimental results revealed that all the evaluated JIT defect prediction models exhibit some level of miscalibration across all these metrics. Surprisingly, even simpler techniques, such as LApredict -which uses logistic regression- show susceptibility to miscalibration. 

These findings emphasize the importance of assessing and, if needed, correcting model calibration before using prediction probabilities as confidence scores to support decision-making in quality assurance teams. Relying on uncalibrated probabilities as confidence scores can mislead practitioners about which predictions to trust, potentially resulting in ineffective prioritization of predictions and inefficient allocation of quality assurance resources.

Motivated by findings, we further experimented with two state-of-the art post-calibration methods. The goal was to assess whether, and to what extent, post-calibration methods can improve the calibration of already trained JIT defect prediction models.
Our experiments showed that, while, in some cases, Platt scaling did significantly reduce miscalibration error (ECE, MCE, and Brier score),  this method introduces complications that require further investigation.
Temperature scaling, on the other side, causes only minor calibration deviations compared to the calibration of the original, uncalibrated models.

In our future work, we intend to enhance the generalizability of this study by expanding our assessment to include other just-in-time defect prediction techniques. Moreover, we intend to broaden our analysis by evaluating the JIT defect prediction techniques evaluated in this study using additional, more recent datasets, such as the datasets collected by Keshavarz et al. \cite{b40} and Ni et al. \cite{b41}. In addition, we plan to investigate alternative calibration methods, such as design time calibration methods. Design-time calibration methods involve integrating the calibration process directly into model training, such that during the training the model learns not only to make predictions but also to adjust its prediction probabilities. However, these methods, typically require the modification of the predictive model architecture or training.

\section*{data availability}

To align with open science principles and support the replication and extension of our study, we have made our study artifacts publicly available. This includes our implementation details (datasets and code) as well as experimental results, all of which can be accessed in the following repository \footnote{\url{https://github.com/An0nym0u5Ssubmi55i0n/On-the-calibration-of-Just-in-time-Defect-Prediction}} \cite{b38}.

\section*{Acknowledgments}
We sincerely thank our colleagues Benjamin Low, Maike Bendel, and Paul-Andrei Dragan for their valuable contributions, insightful discussions, and support throughout this research. We also extend our gratitude to the anonymous reviewers for their constructive comments and suggestions, which have helped us enhance the quality of this paper.

\vspace{12pt}


\begin{thebibliography}{00}

%\bibitem{b0} T. J. Ostrand, E. J. Weyuker, and R. M. Bell, "Predicting the location and number of faults in large software systems," in IEEE Transactions on Software Engineering, vol. 31, pp.340-355, 2005.

\bibitem{b1} Y. Zhao, K. Damevski, and H. Chen, "A Systematic Survey of Just-in-Time Software Defect Prediction," in ACM Comput. Surveys, vol. 55, pp.1-35, October 2023. 

\bibitem{b2} G. Giray, K. E. Bennin, O. Köksal, O. Babur and B. Tekinerdogan, "On the use of deep learning in software defect prediction," in Journal of Systems and Software, vol.195, pp.111537,  January 2023. 

\bibitem{b3} A. Mockus, D. M. Weiss, "Predicting risk of software changes," in Bell Labs Technical Journal, vol. 5, no. 2, pp. 169-180, April-June 2000.

\bibitem{b4} S. Kim, E. J. Whitehead and Y. Zhang, "Classifying Software Changes: Clean or Buggy?" in IEEE Transactions on Software Engineering, vol. 34, no. 2, pp. 181-196, March-April 2008.

\bibitem{b5} X. Wan, Z. Zheng, F. Qin, X. Lu and K. Qiu, "Adjusted Trust Score: A Novel Approach for Estimating the Trustworthiness of Software Defect Prediction Models," in IEEE Transactions on Reliability, pp.1-15,  2024.

\bibitem{b6} X. Shahini, A. Metzger and K. Pohl, "An Empirical Study on Just-in-time Conformal Defect Prediction," in Proceedings of the 21st International Conference on Mining Software Repositories, pp. 88-99, 2024.

\bibitem{b7} C. Pornprasit and C. K. Tantithamthavorn, "Jitline: A simpler, better, faster, finer-grained just-in-time defect prediction," in Proceedings of the 18th International Conference on Mining Software Repositories, pp. 369-379, 2021.

\bibitem{b8} Y. Al-Smadi, M. Eshtay, A.  Al-Qerem, S. Nashwan, O. Ouda, and A. A. Abd El-Aziz, "Reliable prediction of software defects using Shapley interpretable machine learning models," in Egyptian Informatics Journal 24, no. 3, 2023.

\bibitem{b9} D.  Mesquita, L. Rocha, J. Gomes, A. Neto, “Classification with reject option for software defect prediction,” in Applied Soft Computing, vol. 49, pp.1085-1093, 2016.

\bibitem{b10} E. Arisholm, L. C. Briand, and B. J. Eivind, "A systematic and comprehensive investigation of methods to build and evaluate fault prediction models," in Journal of Systems and Software 83, no. 1, pp. 2-17, 2010.

\bibitem{b11} T. Jiang, L. Tan and S. Kim, "Personalized defect prediction," in 28th IEEE/ACM International Conference on Automated Software Engineering, pp. 279-289,  2013.

\bibitem{b12} F. Rahman and P. Devanbu, "How, and why, process metrics are better," in Proceedings of the International Conference on Software Engineering, IEEE Press, pp.432–441, 2013.

\bibitem{b13} Y. Kamei, E. Shihab, B. Adams, A. E. Hassan, A. Mockus, A. Sinha, N. Ubayashi , "A large-scale empirical study of just-in-time quality assurance," in IEEE Transactions on Software Engineering, vol. 39, no. 6, pp. 757-773, June 2013.

\bibitem{b14} E. Arisholm and L. C. Briand, "Predicting fault-prone components in a java legacy system," in Proceedings of the 2006 ACM/IEEE International Symposium on Empirical Software Engineering, pp. 8–17,2006.

\bibitem{b15} T. Mende and R. Koschke,"Effort-Aware Defect Prediction Models," in the European Conference on Software Maintenance and Reengineering, pp.107-116, 2010.

\bibitem{b16} X. Chen, Y. Zhao, Q. Wang, Z. Yuan, "MULTI: Multi-objective effort-aware just-in-time software defect prediction," in Information and Software Technology, Vol. 93, pp.1-13, 2018.

\bibitem{b17} C. Guo, P. Geoff, S.Yu and W. Q. Kilian, "On calibration of modern neural networks," in International conference on machine learning, pp. 1321-1330, 2017.

\bibitem{b18} M. Minderer, J. Djolonga, R. Romijnders, F. Hubis,X. Zhai, N. Houlsby, D. Tran and M. Lucic, "Revisiting the calibration of modern neural networks," in Advances in Neural Information Processing Systems, vol. 34, pp. 15682-15694, 2021

\bibitem{b19} A. Niculescu-Mizil and C. Rich, "Predicting good probabilities with supervised learning," in Proceedings of the 22nd international conference on Machine learning, pp. 625-632. 2005.

\bibitem{b20} R. Caruana and A. Niculescu-Mizil, "An empirical comparison of supervised learning algorithms," in Proceedings of the 23rd international conference on Machine learning, pp. 161-168. 2006.

\bibitem{b21} J. Nixon, W. D. Michael, Z. Linchuan, J. Ghassen, and T. Dustin, "Measuring Calibration in Deep Learning," in CVPR workshops, vol. 2, no. 7. 2019.

\bibitem{b22} J. Platt "Probabilistic outputs for support vector machines and comparisons to regularized likelihood methods," in Advances in large margin classifiers 10, no. 3, pp.61-74, 1999.

\bibitem{b23} B. Zadrozny and E. Charles, "Transforming classifier scores into accurate multiclass probability estimates," in Proceedings of the eighth ACM SIGKDD international conference on Knowledge discovery and data mining, pp. 694-699, 2002.

\bibitem{b24} T. Hoang, H. K. Dam, Y. Kamei, D. Lo, and N. Ubayashi, "Deepjit: an end-to-end deep learning framework for just-in-time defect prediction," in 16th International Conference on Mining Software Repositories (MSR), pp. 34-45. IEEE, 2019.

\bibitem{b25} Z. Zeng, Y. Zhang, H. Zhang, and L. Zhang, "Deep just-in-time defect prediction: how far are we?," in Proceedings of the 30th ACM SIGSOFT International Symposium on Software Testing and Analysis, pp. 427-438, 2021.

\bibitem{b26} X: Zhou, D. Han, and D. Lo, "Assessing generalizability of codebert," in the International Conference on Software Maintenance and Evolution (ICSME), pp. 425-436. IEEE, 2021.

\bibitem{b27} S. McIntosh and Y. Kamei, "Are Fix-Inducing Changes a Moving Target? A Longitudinal Case Study of Just-In-Time Defect Prediction," in Transactions on Software Engineering, 2017.

\bibitem{b28} R. Vasilev and A. D'yakonov, "Calibration of Neural Networks," in ArXiv, 2023.

\bibitem{b29} D. Hendrycks,and K. Gimpel, "A Baseline for Detecting Misclassified and Out-of-Distribution Examples in Neural Networks," in International Conference on Learning Representations, 2022.

\bibitem{b30} X. Jiang X, M. Osl, J. Kim J, and L. Ohno-Machado, "Calibrating predictive model estimates to support personalized medicine," in J. Am Med Inform Assoc, vol.2, PP.263-74, 2012.

\bibitem{b31} A. Kramer and J. Zimmerman, "Assessing the Calibration of Mortality Benchmarks in Critical Care," in Critical care medicine, vol.35. 2007. 

\bibitem{b32} T. Ayer, O. Alagoz, J. Chhatwal, J. Shavlik, C. Kahn, E. Jr and E. Burnside, "Breast cancer risk estimation with artificial neural networks revisited: discrimination and calibration," in  Cancer, 116(14), pp.3310–3321, 2010. 

\bibitem{b33} T. Zimmermann, R. Premraj, and A. Zeller, " Predicting Defects for Eclipse, " in Proceedings of the Third International Workshop on Predictor Models in Software Engineering, 2007.

\bibitem{b34} M. Christakis, and C. Bird, "What developers want and need from program analysis: an empirical study," in  International Conference on Automated Software Engineering, pp. 332-343, 2016.

\bibitem{b35} B. W. Boehm, and P. Papaccio, “Understanding and Controlling Software Costs,” in IEEE Trans. Software Eng., pp.1462-1477, 1988.

\bibitem{b36} T. Gneiting, and M. Katzfuss, "Probabilistic Forecasting," in Annual Review of Statistics and Its Application, 2014.

\bibitem{b37} G. Hinton, "Distilling the Knowledge in a Neural Network," 2015.

\bibitem{b38} Replication package for this study: \url{https://github.com/An0nym0u5Ssubmi55i0n/On-the-calibration-of-Just-in-time-Defect-Prediction}

\bibitem{b39} X. Jiang, and M. Osl, and J. Kim, and L. OhnoMachado, "Calibrating predictive model estimates to support personalized medicine," in Journal of the American Medical Informatics Association, vol.19, pp.263–274, 2012.

\bibitem{b40} K., Hossein, and M. Nagappan, "Apachejit: a large dataset for just-in-time defect prediction," in Proceedings of the 19th international conference on mining software repositories, 2022.

\bibitem{b41} N, Chao, and X. Xu, and K. Yang, and D. Lo, "Boosting Just-in-Time Defect Prediction with Specific Features of C/C++ Programming Languages in Code Changes," in 2023 IEEE/ACM 20th International Conference on Mining Software Repositories, pp. 472-484. IEEE, 2023.






\end{thebibliography}
\end{document}